# How universal is the law of income distribution? Cross country comparison

Ivan O. Kitov[1] and Oleg I. Kitov[2]

[1] The Institute of Geospheres Dynamics, Russian Academy of Sciences
[2] The University of Oxford

Abstract
The evolution of personal income distribution (PID) in four countries: Canada, New Zealand, the UK, and the USA follows a unique trajectory. We have revealed precise match in the shape of two age-dependent features of the PID: mean income and the portion of people with the highest incomes (2 to 5% of the working age population). Because of the U.S. economic superiority, as expressed by real GDP per head, the curves of mean income and the portion of rich people currently observed in three chasing countries one-to-one reproduce the curves measured in the USA 15 to 25 years before. This result of cross country comparison implies that the driving force behind the PID evolution is the same in four studied countries. Our parsimonious microeconomic model, which links the change in PID only with one exogenous parameter - real GDP per capita, accurately predicts all studied features for the U.S. This study proves that our quantitative model, based on one first-order differential equation, is universal. For example, new observations in Canada, New Zealand, and the UK confirm our previous finding that the age of maximum mean income is defined by the root-square dependence on real GDP per capita.

## Introduction

We have carried out a cross country comparison based on two specific characteristics related to personal income distribution (PID). Our study is constrained to data provided by the USA, the UK, Canada, and New Zealand because other countries of interest do not grant open access to income estimates obtained from surveys, tax and administrative records. The U.S. has one of the best systems of income measurements and the longest historic time series, with almost all data (except may be data in the highest income range) available in digital format from various governmental agencies, private companies, and universities. The length, availability, and accuracy of income data make the U.S. a natural reference case for cross-country comparison and a benchmark for relative accuracy estimates.

Our current study is facilitated by the fact that we have already carried out an extensive and detailed study of a diversity of characteristics related to personal income distribution in the U.S. and presented select results [1,2]. Two characteristics were revealed, which best express the evolution of personal income distribution (PID) with age and time. First characteristic is related to the mean income dependence on age [3]. The shape of the corresponding curve evolves in time following up the increasing real GDP per capita. We have found that the age of maximum mean income grows as the square root of the real GDP per capita and this phenomenon is clearly observed in the personal income data published by the U.S. Census Bureau since 1947. In order to quantitatively predict the observed change in the measured PIDs, we have developed a microeconomic model [1], which describes the change in personal income distribution in the USA from the start of measurements. In our model, the mean income dependence on age has one defining parameter – real GDP per capita.

Second studied characteristic is related to the mean income evolution, but it is more sensitive to age. This is the dependence on age of the portion of people with the highest incomes [4]. The dynamics of personal income distribution has a clear bifurcation point near the threshold defined by the transition to the Pareto law. The power law distribution of the highest incomes likely expresses the outcome of numerous stochastic processes as it is observed in numerous natural processes, *e.g.* the frequency distribution of earthquake sizes. The driving forces behind these processes are hard to indentify and control, but the net result of their joint work is easy to predict quantitatively. Our model proves that in the lower-income range the evolution of personal income is well defined by a simple differential equation [2]. Among other measured features, this equation accurately predicts the portion of people reaching the Pareto threshold and describes its evolution with real GDP per capita. Therefore, we have a useful tool to model two involved characteristics. In the current study, we prove the universal character of PID evolution as a unique function of GDP, and thus confirm the consistency of our model, by extending the set of observations to three other countries.

In both studied variables, the level of income aggregation over population balances the clarity/reliability of observed changes and the level of suppression of larger fluctuations, which are related to data quality, i.e. to numerous revisions to survey questionnaires (e.g. revisions to personal income definition) and the accuracy of measurements themselves. As reported by the Census Bureau, the share of population with income in the total working age population increases from 73% in 1960 to 91% in 2000, chiefly because of definition of personal income adopted in the CPS. Two studied characteristics demonstrate measurable changes since 1962, but the estimates of Gini ratio, obtained from the same Current Population Survey (CPS) personal incomes, show just marginal fluctuation between 0.50 and 0.52 [2]. We have used the CPS data for people with reported income to calculate a consistent time series for Gini ratio for the whole period between 1947 and 2014. At the same time, fluctuations observed in the original microdata are so high that there is no explanation of the observed changes in narrow age and income intervals for adjacent years except those related to the measurement accuracy. Aggregation in wider income and age bins irons these fluctuations out.

The age dependence of mean income and proportion of people with the highest incomes provides reliable measures of the evolutionary behavior of income distribution as well as robust statistical inferences related to the driving force behind this evolution. In the historical prospective, the U.S. dataset provides likely the most accurate and longest time series describing time and age dependent processes in the PIDs. However, there is always a question about the universal character of the observed evolution with real GDP per capita. This study answers this question to the extent limited by the availability and accuracy of personal income data. (We are going to proceed with new dataset when available.) On the whole, we give a positive answer – the dependence of mean income and the proportion of people with the highest incomes on age as observed in the USA is reproduced one-to-one (considering data accuracy) in Canada, New Zealand, and the UK. The corresponding curves observed in these three countries repeat those observed in the U.S. for the years when the level real GDP per capita was the same. This observation seems to prove that the evolution of personal income (at least in these four countries) follows the same path, i.e. this is a universal characteristic related the only driving force - real GDP per capita.

## 1. Personal income: definitional issues

We start with a slightly provocative statement that there exists no comprehensive and accurate definition of personal income, which can be used for a true estimate of age/gender/race-dependent properties. There are several operational definitions of personal income measuring different portions of the true personal income, the definition of which does not exist so far. Having no genuine income values one is forced to use only available data. In such a situation, the accuracy and coherency of sequential estimates are two most important issues. To conduct a reliable quantitative analysis, one can use any constant portion of the true value and get almost the same predictive power of the obtained relationships as that obtained from the true value itself. For example, a voltmeter accurately measures a voltage using just a small portion of total electric current.

When several sources of data are available it is worth to compare how similar are the features we study as estimated from different datasets. For example, do they reveal the same dependence of mean income on age? This is important aspect of the study by itself since it provides a reasonable constraint on data accuracy. In a cross country comparison, it is especially important because the datasets reported by countries may have different sources. For example, the UK provides an extensive set of income time series based on tax data, while New Zealand publishes the results of annual income surveys conducted in the second quarter of each year. For the U.S. both sets are available and thus it is a straightforward task to compare them before we study other countries.

Figure 1 schematically compares the personal income estimates as reported by three agencies, which provide related statistics: the Bureau of Economic Analysis (BEA), the Census Bureau (CB), and the Internal Revenue Service (IRS). The BEA has been reporting accurate aggregate estimates of the total personal income (PI) and its distribution over major sources (e.g., wage and salary, contributions for employee pension and insurance funds, personal income receipts on assets, etc.) since 1929. With all important procedures, tools and estimates allowing tight control over other sources of income data, the BEA does not provide fine internal structure of the personal income distribution in the USA. For the purposes of our study, it does not report age, gender and race distribution of personal incomes. This set of personal data cannot be used for modelling and comparison of age dependence. In Figure 1, only the ratio of total personal income (PI) and Gross Domestic Income (GDI=GDP) since 1947 is presented. In the year of 1947, the annual income surveys in the U.S. were started.

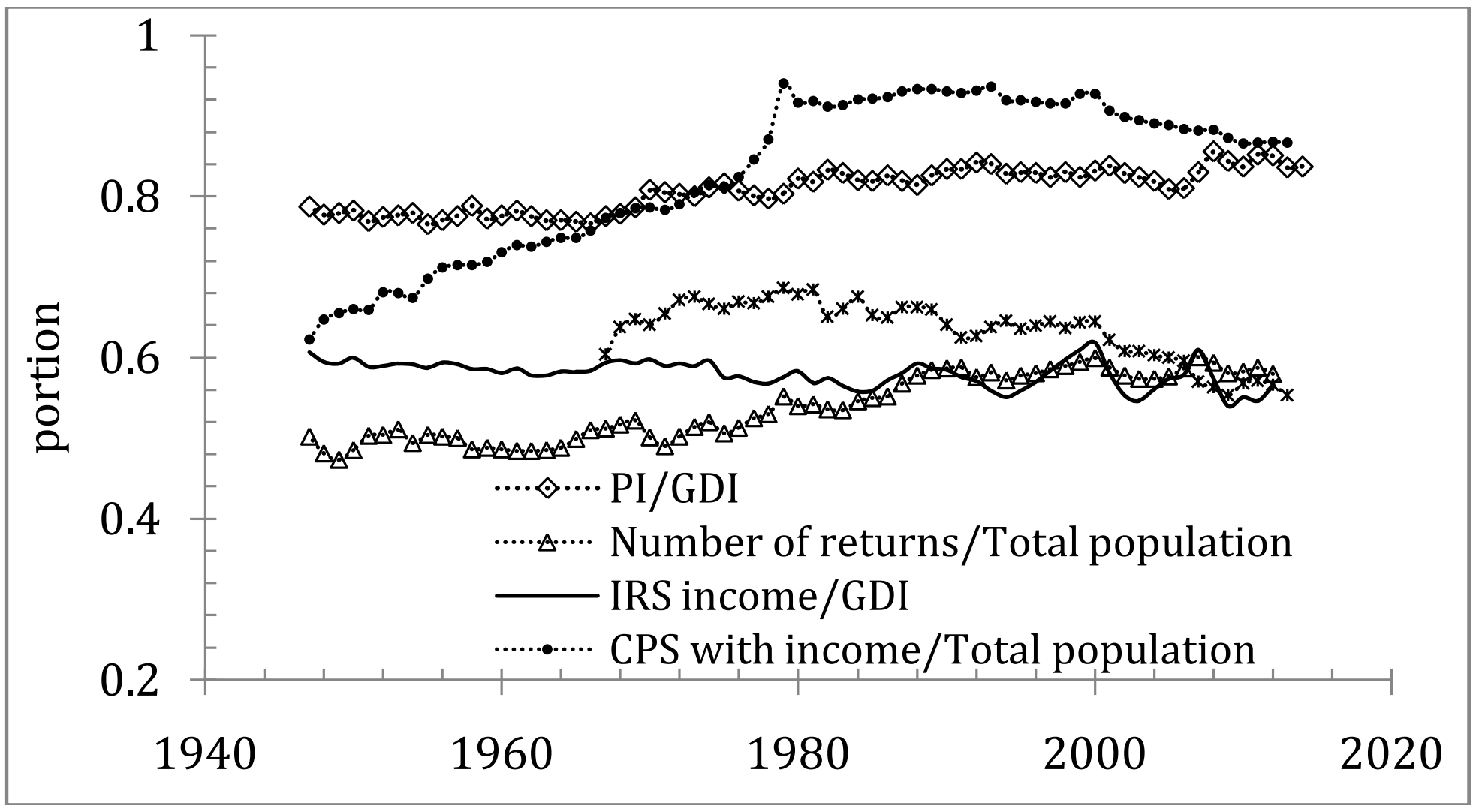

Figure 1. The portion of people with income reported by the Census Bureau from the Current Population Surveys (CPS) and the number of returns reported by the Internal revenue Service (IRS) in the total working age population. The proportion of total personal income reported by the CB and IRS in the Gross Domestic Income (GDI=GDP). For comparison of various definitions, the estimate of personal income estimate (PI) reported by the Bureau of Economic Analysis is shown.

The Internal Revenue Service provides the longest time series of income data – some variables begin in 1913. Figure 1 depicts two time series related to the IRS. The number of individual tax returns (a proxy to the number of people with income) is divided by the total working age population (age 15 and above) for the same year and represents the portion of people with income. In the 1950s and 1960s, the proportion of people with (IRS) income was between 50% and 52%. In the 2000s and 2010s, this proportion was between 57% and 60%. The total income reported to the IRS, which is called Adjusted Gross Income (AGI) is divided by the GDI and represents the portion of personal income (according to the IRS definition) in the GDI. The BEA carries out annual inspections of all incomes included in the AGI and reports very specific errors in the IRS statistics. For example, the gap between the AGI estimate reported by the NIPA (National Income and Product Accounts) and that of the IRS reached 15% in 2005. This gap puts some constraint on the accuracy of the IRS personal income estimates. Unfortunately, these errors are aggregated. They are not distributed over age and so on.

There are several income time series reported by the IRS, which can be used in our study. They include the number of people in finite bins extended to $10,000,000. This is a very high income in comparison with the current level of $250,000 in the CB data, which is, however, reported only from the mid-2000s. The IRS datasets are of crucial importance for estimating the properties of the highest incomes distribution, i.e. the portion of people above some high threshold. The Pareto law implies that the PID above such a threshold should follow up a power law. In addition, the income distribution in five-year age bins (and for two genders) is published for the year of 1998. Similar distributions for ten-year bins are available for the years between 2008 and 2012.

The Census Bureau provides the finest distributions of personal income over numerous parameters. The CB uses the mechanism of annual Current Population Surveys (CPS) to measure personal incomes in approximately 80,000 households. In order to project this smaller population subset to the entire U.S. population the CB uses the age-gender-race dependent scaling coefficients for each person in the CPS. By construction, this approach has much lower measurement accuracy for poorly represented categories. For example, young black females with higher incomes have very low probability to be present in the CPS. Sometimes, one or two persons represent the whole population in the same age-gender-race category. As a consequence, larger fluctuations are observed in the related income distributions.

In Figure 1, the portion of population with income (as reported in the CPS) in the total working age population is presented together with the portion of the CPS total income in the GDI. The total CPS income is also the estimate obtained from the 80,000 households and then scaled to the whole population. In 2012, the CPS population universe presumably included approximately 87% people with income; the IRS gave only 57%. At the same time, the figure of total personal income reported by the CB was about 57% of the GDI, i.e. the same as reported by the IRS. The CB has a quite specific definition of personal income: "CPS money income is defined as total pre-tax cash income earned by persons, excluding certain lump sum payments and excluding capital gains", while "BEA personal income is the income received by persons from participation in production, from government and business transfer payments, and from government interest. BEA estimates personal income largely from administrative data sources" [5]. The CB's personal income estimates are also known as "CPS money income". It is important that larger part of the difference between the CB's and BEA's estimates (around 25% of the CB's income) can be explained by the differences in income sources. The error in wage and salary estimates (underreporting) in the CPS can reach 5% to 10% of the total CPS income.

This is the level of accuracy of personal income data we have to work with. In addition, the IRS dataset has a significant problem with the population coverage. The portion of people with income not only relatively small but also varies with time. Together with the observed high-amplitude oscillations in the AGI, the variations in the number of returns induces measureable fluctuations in the high-level estimates of inequality (e.g. the Gini ratio), with the less aggregated measures of personal income experiencing even larger disturbances. In this sense, the CPS data set has clear advantages since all income estimates are scaled to the whole population. As a consequence, the measures of income inequality are not changing much due to data inconsistency over time.

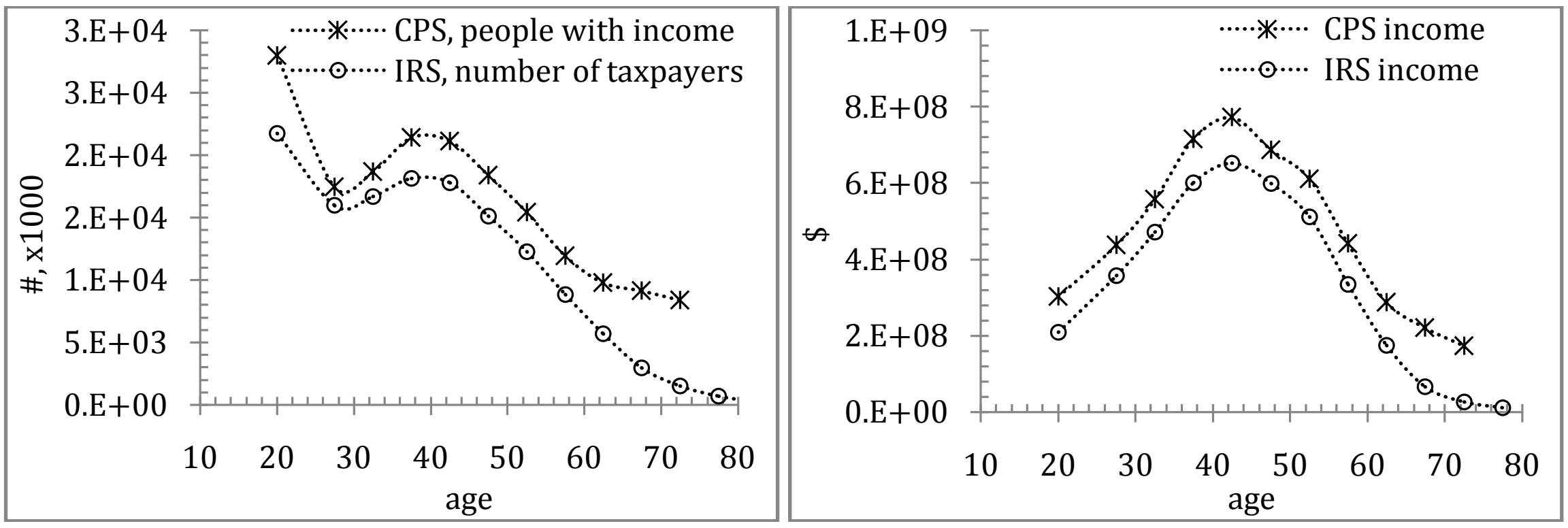

Figure 2. The number of people (left panel) and total income (right panel) in 5-year bins (except the bin between 15 and 24 years of age) as reported by the IRS and CPS for the year of 1998.

Detailed comparison of the PIDs reported by the IRS and CB is beyond the scope of this study. Merging of the IRS high-income data and CPS low-income estimates is a delicate issue and deserves a special study, which may result is a more reliable time series based on a more precise definition of personal income. Here, we compare only one of the characteristics under investigation – the evolution of mean income with age. Figure 2 displays the distribution of people and income over age as reported by the CPS and IRS for the year of 1998. This is the only year when the age dependent mean income is reported by the IRS in narrow age bins. For the years between 2008 and 2012, the IRS reported mean incomes in age bins incompatible with those used by the CB. The CPS counts more people in all age bins and larger total incomes everywhere.

Dividing the total income by the number of people in a given age bin we obtain an estimate of mean income. Figure 3 depicts various mean income curves as reported by the IRS and Census Bureau. To present the CPS data, we use three different data sets. The IPUMS provides the original income measurements (microdata), which allow calculation of mean incomes in one-year bins. The resulting curve in Figure 3 is characterized by visible fluctuations, which are especially large near the peak mean income. Therefore, these estimates are not helpful for accurate comparison with the IRS data. The mean income reported by the CB in five-year bins (see Table PINC-01. Selected Characteristics of People 15 Years and Over, by Total Money Income in 1998, Work Experience in 1998) provides the best case comparison with the IRS data, while the mean income distribution in ten-year bins may cause some problem in the estimation of the age corresponding to the peak mean income.

Figure 3 illustrates the difference in income sources used by the CPS and IRS. The Census Bureau reports higher mean incomes for the youngest and eldest population. At the same time, the IRS mean income is higher for ages between 40 and 65 years. The largest difference is observed near the peak mean income. Despite the discrepancy in the IRS and CPS curves, the most important observation is that the age when these curves reach their peak values are very close. To better illustrate the coincidence of peaks in the mean income distribution we normalized all curves to their peak values.

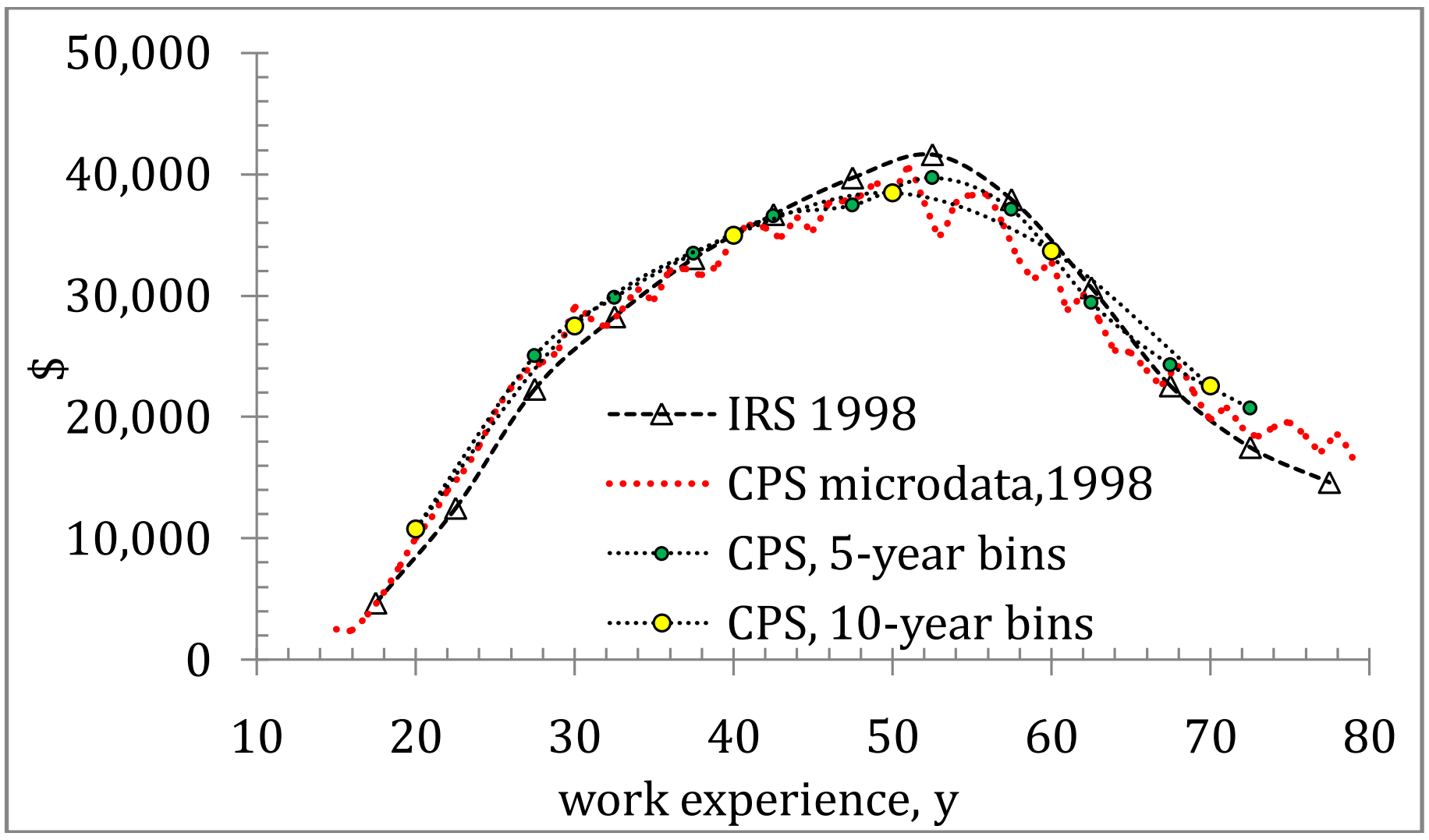

Figure 3. The mean income as a function of age as reported by the Census Bureau (CPS) and the Internal Revenue Service (IRS) for 1998. This is the only year with a fine (five-year age bins) age-dependent personal income distribution reported by the IRS. For comparison, we use the set of microdata published by the IPUMS for 1998. To illustrate the difference in the presentation of income in various bins, the mean income curve based on ten-year bins borrowed from the CB historical dataset is also shown. The mean income for the youngest and eldest age groups is higher than that reported by the IRS. Between 40 and 65 years of age, the IRS mean income is higher.

Figure 4 displays two curves for five-year bins and also presents the curve obtained from the IPUMS microdata but now smoothed with a centred nine-year moving average – MA(9). All three curves are characterized by peak at 52.5 years. This is the age of the largest mean income in 1998. As we know, the peak age in the U.S. has been increasing with real GDP per capita and was above 55 years in 2012.

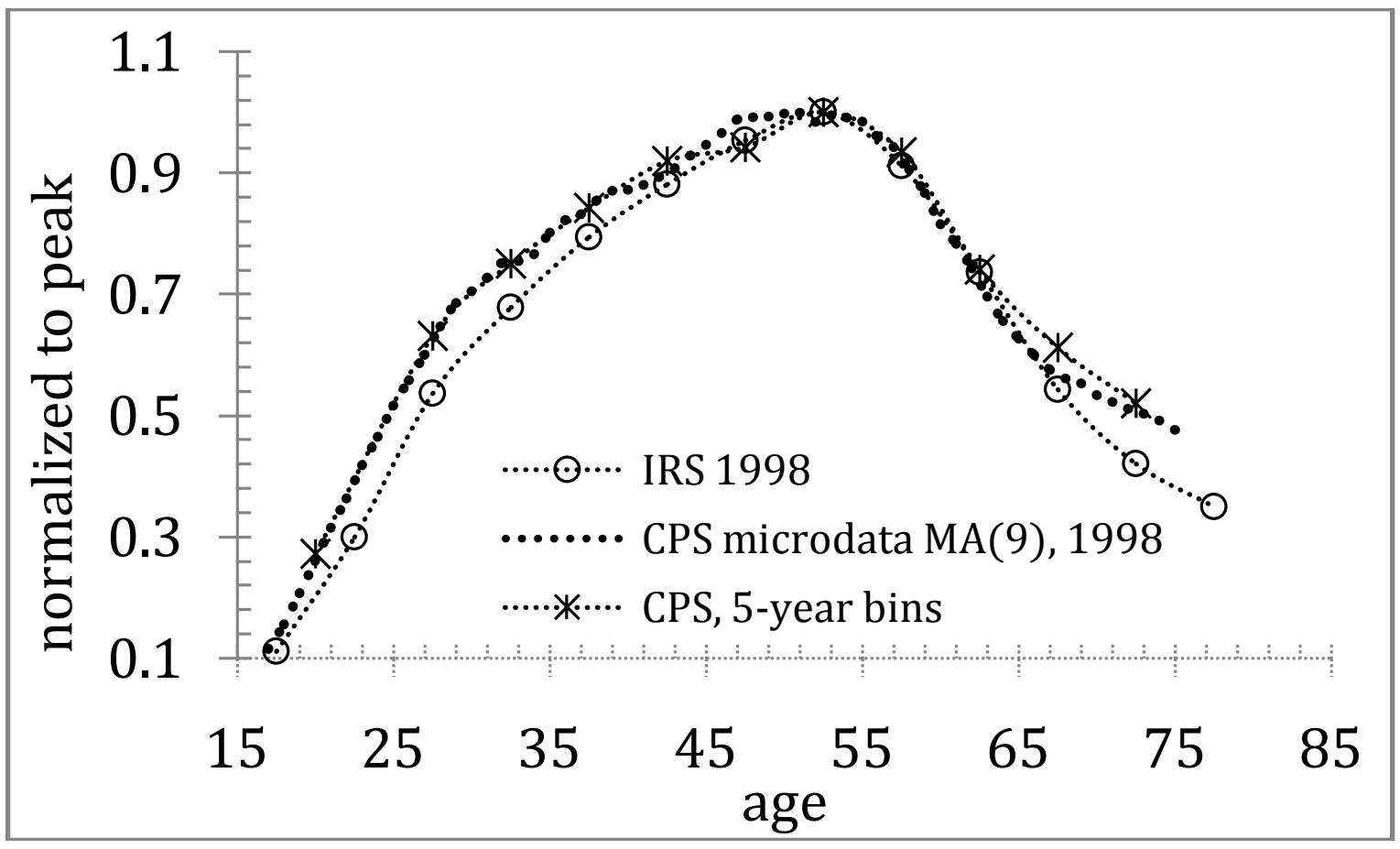

Figure 4. The IRS and CPS microdata curves in Figure 3 are normalized to their peak values. The age of peak mean income is the same for the CPS and IRS.

For the same year, the income estimates provided by the Census Bureau and IRS are close in dimensionless representation. This result is valuable for further comparison of income data obtained by various agencies in different countries. For example, we assume that the similarity of CPS and IRS data in the U.S. allows us to compare tax-related data reported by the UK for a given year and some CPS data twenty to thirty years earlier.

## 2. Cross country comparison

### 2.1. GDP per capita

The main assumption of our study is the dependence of aggregated properties of personal income distribution on real GDP per capita only. This implies that countries with lower GDP per capita have to repeat PID features observed in the countries with higher GDP per capita. To compare different countries we borrowed real GDP per capita estimates from the Total Economy Database (TED) operated by the Conference Board. Figure 5 displays the evolution in the USA, UK, Canada, and New Zealand as expressed in 1990 US$ converted at Geary Khamis PPPs. As we assume that the accuracy of income measurement has been increasing with time the most recent estimates for the latter three countries likely have some advantages to be used for a reliable cross country comparison.

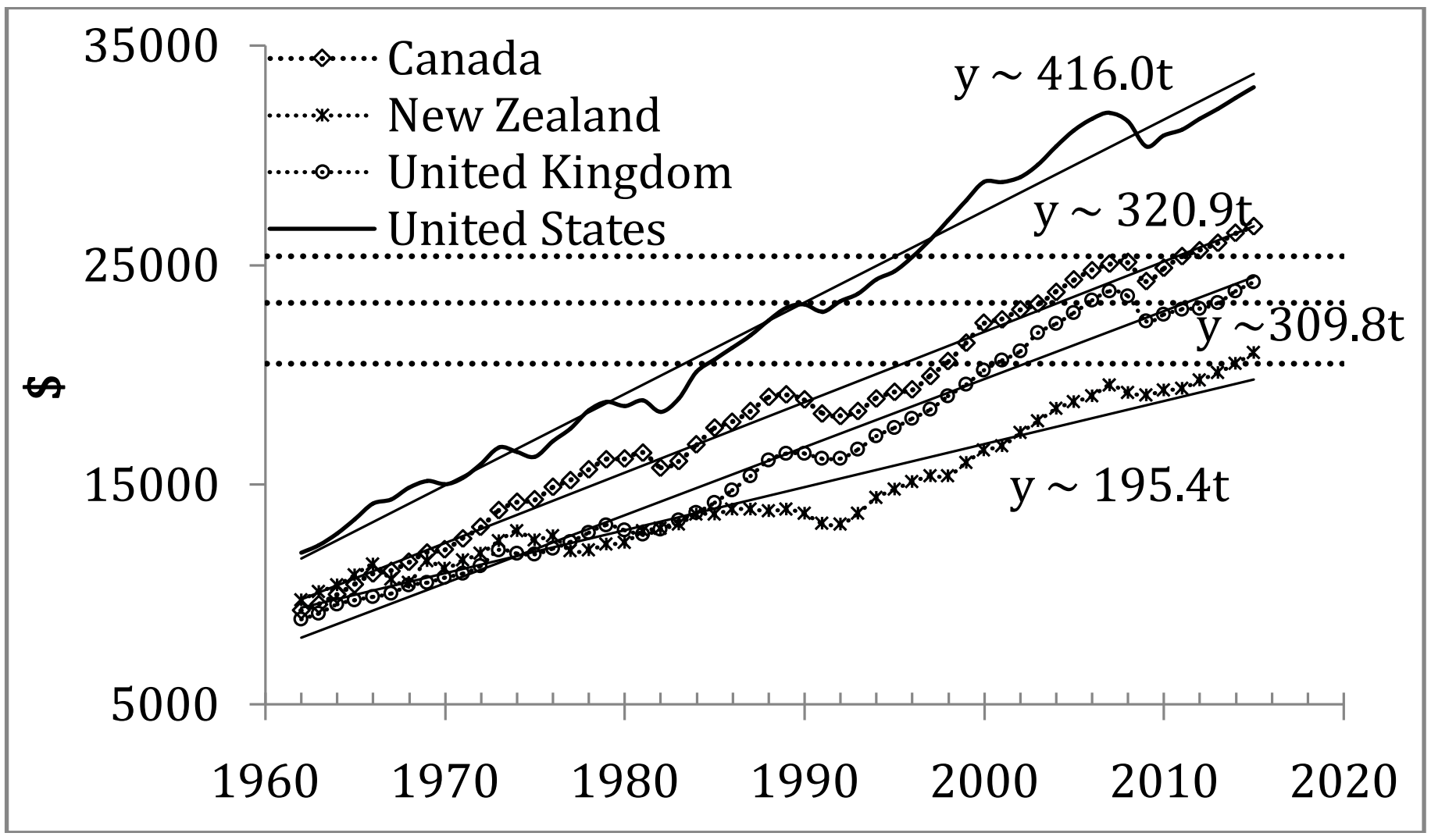

Figure 5. The evolution of real GDP per capita in 1990 US$ (converted at Geary Khamis PPPs) as borrowed from the Total Economy Database reported by the Conference Board. Cross comparison of the dynamics of personal income evolution in four (English speaking) developed countries for which we have retrieved data from open sources – Canada, New Zealand, the UK, and the USA. Three horizontal dotted lines show the level of GDP per capita in three countries in some country-specific years between 2011and 2014 and their intersections with the USA curve provide the years when the US personal income presumably had the same characteristics.

In Figure 5, three horizontal dotted lines show the level of GDP per capita in some country specific year between 2011 and 2014. Their intersections with the USA curve provide the years when the U.S. personal income presumably had the same aggregated characteristics. For example, real GDP per capita in the UK was $23,272 in 2012. In the USA, approximately the same level was observed in 1992 ($23,363). For New Zealand, the estimate of $20,526 in 2014 gives 1985 as the matching year. In Canada, the level of $25,400 in 2011 corresponds to that measured in the USA in 1996. One has to take into account that the PPP values are subject to revision and may not be accurate for some countries and years. Moreover, the PPP curves may differ dramatically from those expressed in domestic currency.
The growth in GDP per head in four countries can be precisely interpolated by a linear function of time [5] with the annual increment (linear regression coefficient) falling from $416 per year in the USA to $195 per year in New Zealand. The obtained linear trends can be extended into the future and provide the estimates of expected time when the level of real GDP per capita in a given country will reach any projected value. For example, the level of $32,000 (1990 US$) observed in the U.S. in 2007 will be reached in Canada in 2031, in the UK – 2040, and only in 2071 in New Zealand.

The GDP estimates reported by various agencies may differ by a few percent. For example, Figure 6 depicts the ratio of BEA (chained 2014 US$) and TED (chained 1990 US$) real GDP per capita estimates (right scale) between 1947 and 2013. Presumably, this ratio has to be constant as expressed by the difference between 2014 and 1990 US$. However, it has been growing from 1.506 in 1947 and 1.545 in 2000. The largest difference is approximately 2.6%. For the USA, such low amplitude effect does not influence the accuracy of the peak age estimate in the dependence of mean income on real GDP per capita. For other countries under consideration, the effect of GDP measurement accuracy may have larger amplitude and should not be neglected when the peak age difference is estimated.

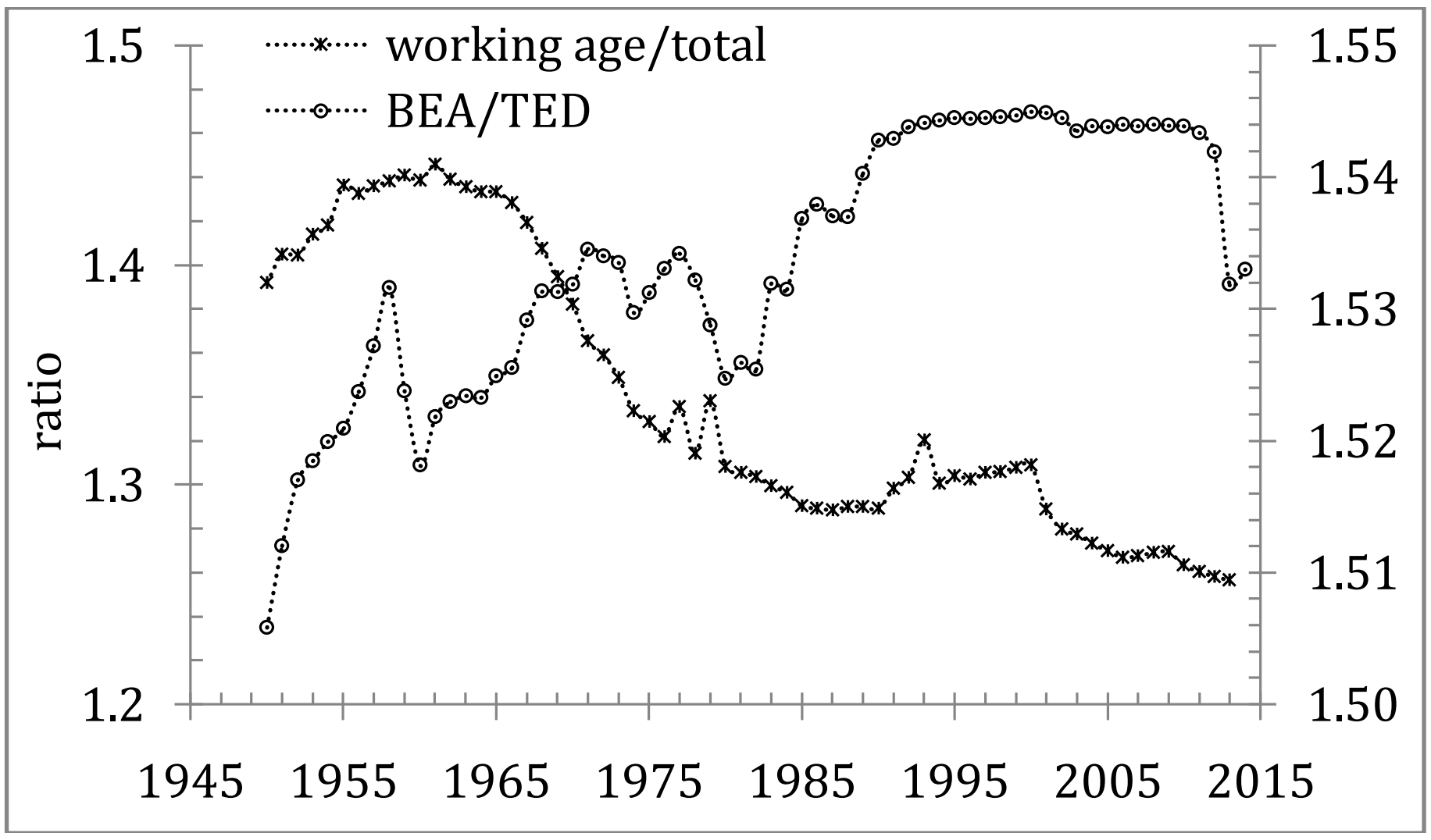

Figure 6. The ratio of total and working age population (left scale) decreases from 1.44 in 1960 to 1.26 in 2013. The correction of real GDP per capita for the difference between total and working age population (income earners) is 0.87, i.e. the change in real GDP per capita is really by a factor of 0.87 smaller than that estimated from the TED or BEA. Same corrections have to be done to all GDP time series in this study. The ratio of BEA (chained 2014 US$) and TED (chained 1990 US$) real GDP per capita estimates (right scale) varies between 1.506 in 1947 and 1.545 in 2000 (right scale). This ratio expresses the relative accuracy of real GDP measurements.

For the purposes of our quantitative analysis we have to use real GDP estimates as the defining parameter. For the USA, both time series (TED and BEA) are estimated as the total GDP divided by total population. Nothing is wrong with this definition often used to illustrate real economic growth, but it includes people below 15 year of age who are never included in income statistics. To obtain an unbiased time series, one has to correct the reported GDP estimates for the ratio of total and working age population. Figure 6 shows that this effect is not marginal at all – this ratio decreases from 1.44 in 1960 to 1.26 in 2013. Therefore, the ratio of real GDP per capita estimates in 2013 and 1960 has to be corrected by a factor of 0.87 – from 2.83 to 2.46. Since we suggest a square root dependence of the peak age in mean income curve on GDP the expected change between 1960 and 2013 has to be 1.57 times instead of 1.68. If the peak age in 1960 is, say, 41 year (27 years of work experience) then the projected age with the TED GDP series is 58.6 years (44.6 years of work experience), while it is 56.4 years with the corrected GDP estimates. The two year difference demonstrates the importance of similar corrections to all GDP time series in this study.

### 2.2. The United Kingdom

The United Kingdom is the first country to compare with the U.S. All income-related tables were borrowed from [6], which is a part of the UK Government portal. The age-dependent income data are obtained from the Survey of Personal Incomes (SPI), an annual sample survey carried out by HM Revenue & Customs. The income tables include only information on individuals liable to UK income tax, i.e. sources of income are restricted to tax purposes only. Therefore, the UK income tables are better to be compared to those reported by the U.S. IRS. In paragraph 2.1, we found that the level real GDP per capita in the UK lags by about 20 years behind that measured in the USA. Under our framework, the shape of the age-dependent mean income curves depends only on real GDP per capita. Then the U.S. curve to compare to the 2012 UK mean income is that for 1992. Since the data on the years before 1993 are not available from the IRS we have to use the CPS income tables. Such a replacement may introduce some distortions in matching process. However, as shown in Section 1, the IRS and CPS estimates of peak mean income age are very close.

We start with the time history of age-dependent mean income in the UK, which is confined to the years between 1999 and 2012, since the original tables (1999-2000 through 2012-2013) are available only for this period. There are no estimates for 2008 as the relevant table is not published. Figure 7 presents the evolution of age-dependent (nominal) mean income (expressed in GBP) with time. The presented curves are obtained as spline interpolations between actual estimates in 5-year age bins. The shape of these curves is similar to that observed in the U.S. – quasi-logarithmic growth to the peak value and then quasi-exponential fall. The age of transition from growth to fall, *i.e*. the age of peak mean income, has been increasing with time. If the shape of the mean income curve depends only on real GDP per capita the age of peak in the UK has to follow up the same trajectory as in the U.S., with the GDP per capita as defining parameter.

For the purpose of quantitative analysis, age is replaced by work experience. By definition, personal work experience is equal to the age of a given person less 14 years. The mean income estimates are assigned to the midpoints of the respective work experience bins, as shown for the 1999 curve in Figure 7. The only exception is the youngest and open-ended population bin “under 20 years of age”, where no midpoint can be assigned. In Figure 7, we assign the corresponding mean income value to 2 years of work experience by force and use this estimate only for illustration. The 2012 curve is above all other curves everywhere. The 1999 curve is the lowermost one except in the large work experience range. Three curves between 2001 and 2003 are very close to each other. Almost all curves are characterized by local fluctuations of varying amplitude in the bin between 45 and 49 years of work experience. The curves between 2005 and 2007 are smooth, however. The cause of this difference is not clear and we do not consider all possible deviations between the UK and U.S. mean income curves in this age range.

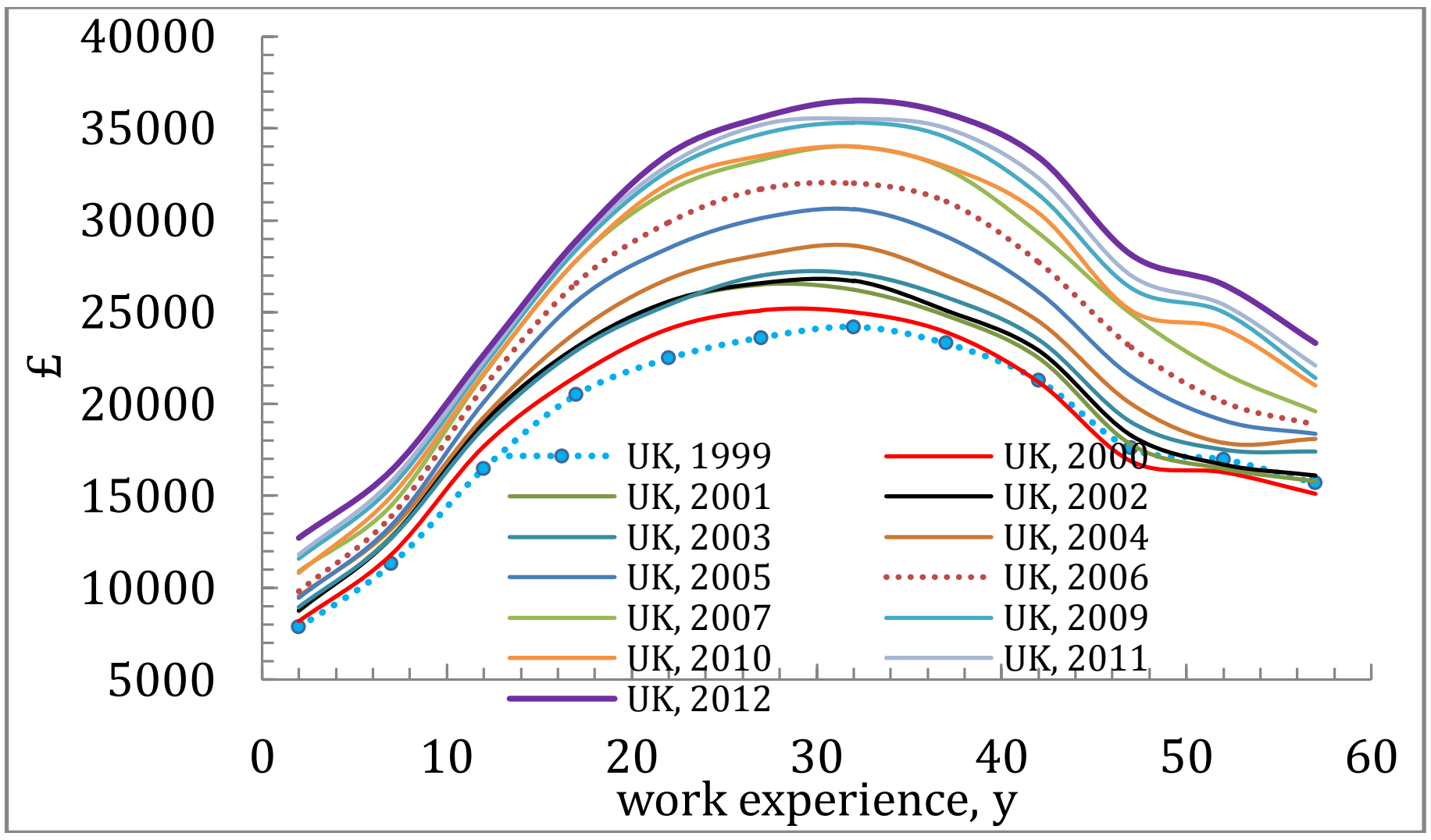

Figure 7. The evolution of age-dependent mean income in the UK between 1999 and 2012. No estimates are available for 2008. The absence is likely related to the 2008 financial crisis. The mean income values are assigned to the midpoints of 5-year bins as shown by circles in the 1999 curve. Age is replaced by work experience. The 2012 curve is above all other curved everywhere. The 1999 curve is the lowermost one except in the large work experience range. Three curves between 2001 and 2003 are very close to each other. Almost all curves are characterized by local fluctuations in the bin between 45 and 49 years of work experience. The curves between 2005 and 2007 are smooth, however.

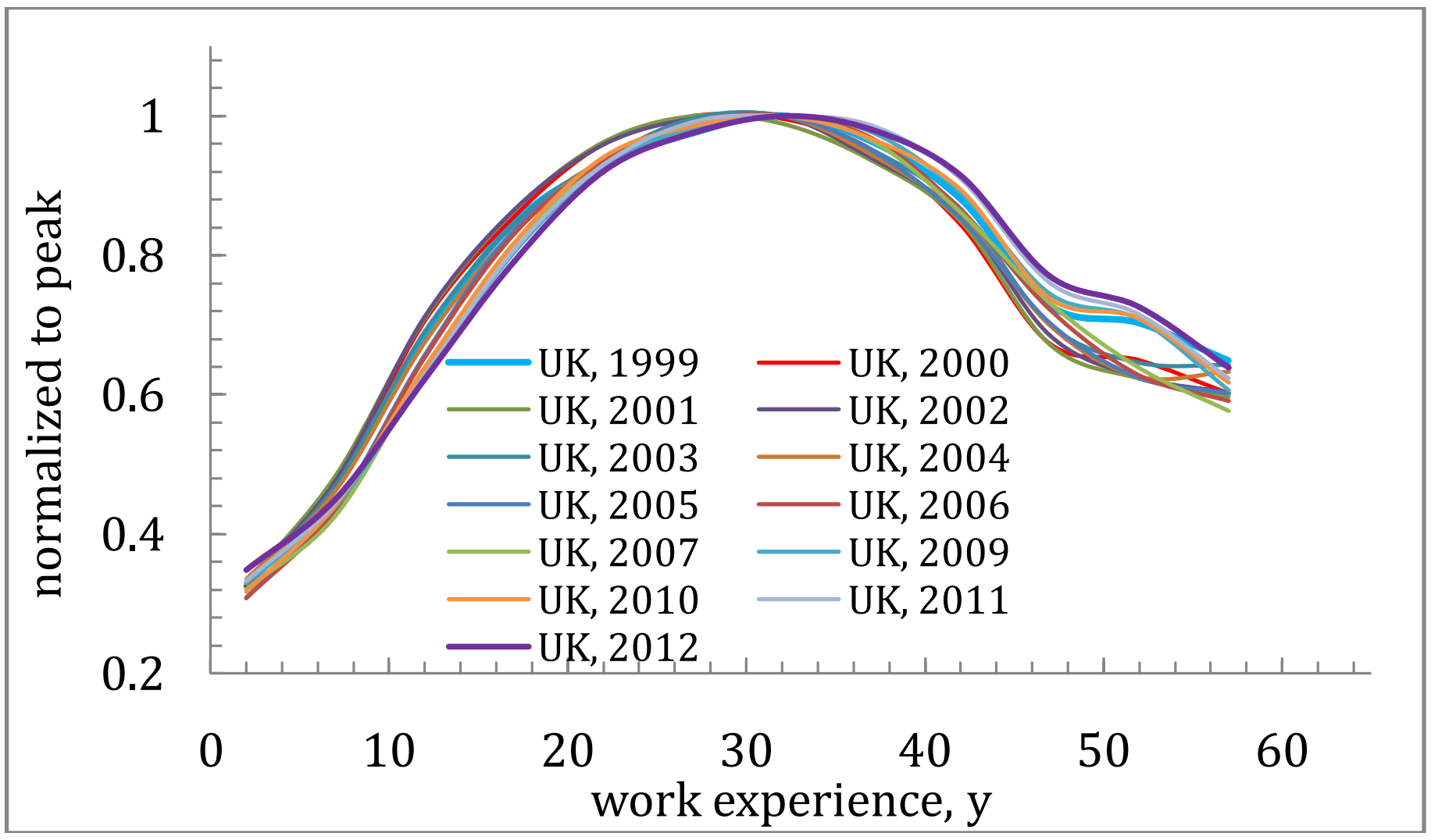

Figure 8. Same curves as in Figure 7 with all mean income estimates normalized to the peak mean incomes for the respective years. The growth in the work experience corresponding to the peak value is seen.

Since individual incomes in different countries are measure in domestic currency one cannot carry out a direct comparison of mean income curves. Scaling to some common currency (e.g. USD) is possible at, say, PPP conversion rates, but in terms of shape comparison this procedure would not differ from the normalization of the mean income curves to their respective maximum values. This was a standard procedure in comparison of U.S. incomes for the period between 1947 and 2011 [2] and we have applied in to the UK data. Figure 8 displays the same curves as in Figure 7 but normalized to their respective peak values. There are two clear observations – the work experience corresponding to the peak mean income increases with time (actually GDP per capita) and the 2012 curve now lies below all other curves before the peak value and above after the peak work experience. We call the age (work experience) corresponding to the peak mean income "critical age" or "bifurcation point" since the behaviour of the mean income curve changes from quasi-saturation growth to exponential fall. In this critical point, the process of income distribution suffers some dramatic changes, and this is not the age of retirement. In the U.S., the critical age was measured between 35 and 40 years of age seventy years ago and currently approaches 60.

In order to estimate the age of peak mean income, Figure 9 presents the central segments of the curves in Figure 8. The normalization procedure results in dimensionless estimates of the average income assigned to the midpoints of the corresponding age bins. The lines drawn through these estimates do not represent actual values of dimensionless mean income except in the midpoints. They are spline interpolations of these estimated values, with the age of the maximum value in the obtained curves likely to be shifted from the midpoint of the bin with the peak value. (Therefore, some curves may be above 1.0.) These estimated maxima are then used to evaluate the shift in the age-dependence mean income. In Figure 9, the 1999 curve (thick blue line) has a larger work experience peak (around 31 years) than the peak age for 2000 and 2001 (dotted lines) - between 28 and 29 years, as well as for 2002 and 2003 (dashed lines) – around 30 years. The curves between 2004 and 2011 are characterized by a gradual increase in the peak age, with the 2012 curve (black line) peaking at approximately 32.5 years of work experience. Therefore, the 1999 curve likely includes some biased estimates and we do not use it in the following quantitative estimates.

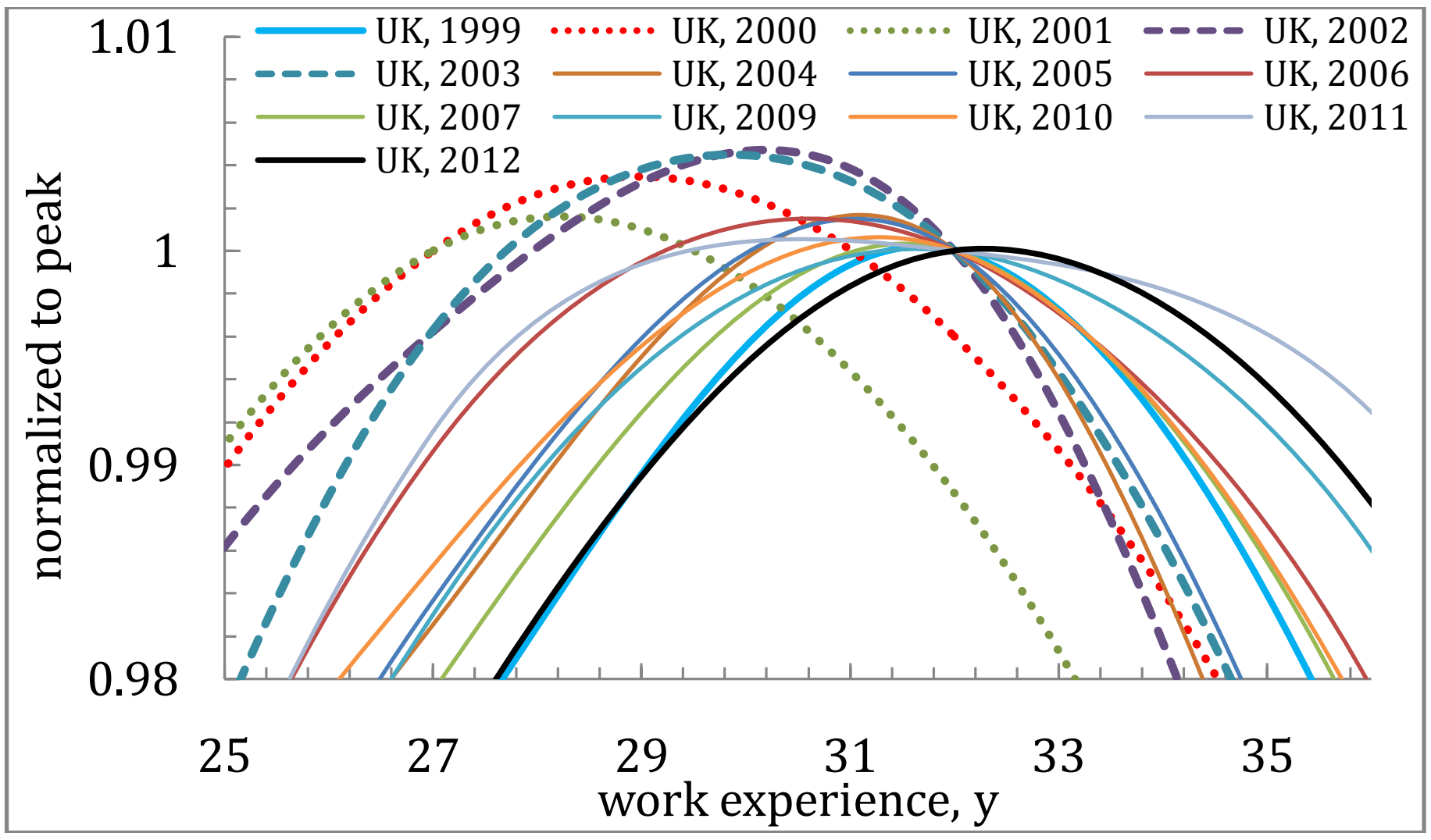

Figure 9. The age (work experience) of peak mean income increases with real GDP per capita. The 1999 curve (thick blue line) has a larger work experience peak than those for 2000 and 2001 (dotted lines) around 28 years, as well as for 2002 and 2003 (dashed lines) – around 30 years. The curves between 2004 and 2011 are characterized by gradual increase in the peak age. The 2012 curve (black line) peaks at 32.5 years.

Figure 9 demonstrates the evolution of the normalized curves and proves that the work experience corresponding to the peak mean income increases from approximately 28-29 years in 2000 and 2001 to above 32 years in 2012-2013. The estimates of real GDP per capita are \$20,207 and \$23,017 in 2000 and 2012, respectively. Theoretically, the working experience should increase from 28.5 years to $28.5\sqrt{(23017/20207)}=30.9$ years. The estimates of average disposable income reported by the OECD give a slightly larger age growth between 2000 and 2012 - 2.7 years. So, the assumed root square dependence suggests that the theoretical difference between the peak ages has to be approximately 2.4 to 2.7 years.

Considering the accuracy of the peak age and GDP/income measurements, the match between the predicted increase of 2 to 3 years and the observed one of approximately 4 years is a good one. For better estimate, one needs a much longer and more accurate time series.

Historically, population in the UK needs more and more time to reach the peak mean income. According to our microeconomic model [2], this effect is caused by increasing sizes of work capital, with the growth proportional to the root square of the real GDP per capita, in line with the Cobb-Douglas production function. This allows higher personal incomes (as well as real GDP per capita) to be achieved by all individuals in a given economy. Basically, the mechanism allowing getting higher incomes consists in decreasing discounting factor counteracting income growth. For a given person, the rate of income discounting is proportional to the attained level of income and inversely proportional to the size of work capital applied by this person. Mathematically, this term leads to a slower relative discounting for incomes earned with the largest work capitals.

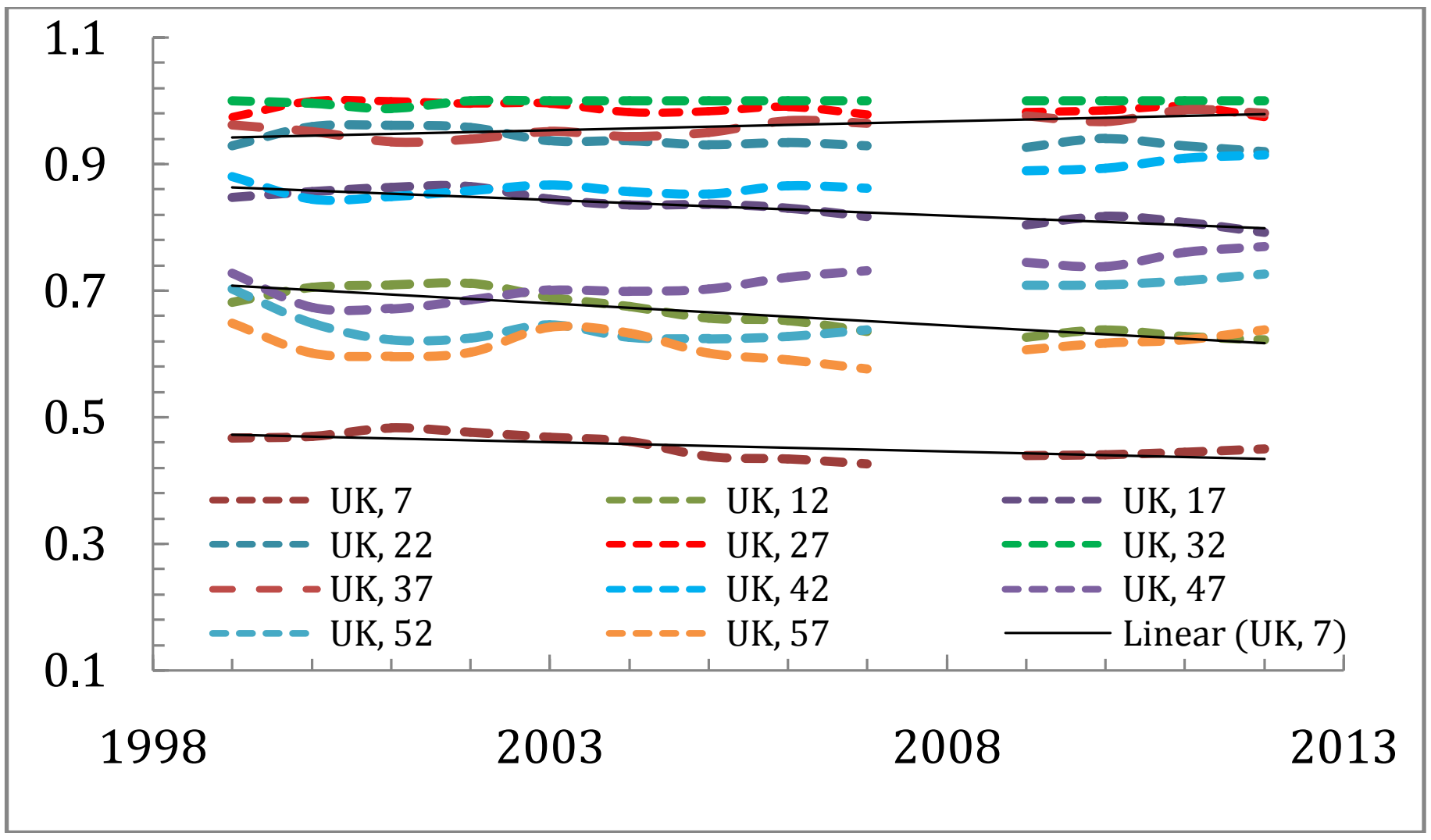

Figure 10. The evolution of mean income in all 5-year-wide age groups normalized to the peak value in the same year. In three youngest age groups (7, 12, and 17 years of work experience), the proportion of mean income has been falling since 1999. The peak work experience shifted from the 25 to 29 years bin to the 30 to 34 years bin between 2000 and 2003. The proportion of mean income has been increasing in the group between 35 and 39 years. The peak will likely move into this age group in the next 10 to 15 years, depending on real GDP growth. The proportion of mean income in the elder categories has been increasing as well.

One negative outcome of the increasing real GDP per capita is that the relative share of income in the youngest age group is subject to gradual decrease, which is inevitable in the current system of economic and social ties. Figure 10 presents the evolution of the normalized mean income in all age groups; the group “under 20” is not shown. Three youngest age groups: 5 to 9, 10 to 14, and 15 to 19 years of work experience, are characterized by a falling proportion of their mean income since 1999. This trend cannot be reversed in the future if real GDP per capita will be growing. Between 2000 and 2003, the peak work experience shifted from the 25 to 29 years bin to the 30 to 34 years bin. Such a transition happens not often and we are lucky to find it in the UK data taking into account the shortness of the time series. The next transition will be to the 35-39 years bin. One can observe that the proportion of mean income has been also increasing in this group. The peak mean income will likely move into this age group in the next 10 to 15 years, depending on real GDP growth. The portion of mean income in the elder categories has been increasing as well. The 25-29 years of work experience group has lost the peak and joined the category of younger population.

The increasing age of mean income peak and the decreasing income portion of the youngest population, as observed in the UK, both confirm similar features measured in the mean income distribution in the USA. Therefore, the evolution of personal income distribution in the UK likely follows the same dependence on age as in the USA. As discussed above, in case this dependence is, mathematically, a universal one, the curves in Figure 8 have to repeat similar curves observed in the U.S. in the years defined by GDP per capita.

In paragraph 2.1, we have evaluated that the UK curve for 2012-2013 has to match the U.S. curve for 1992. For this comparison, we used income microdata published by the IPUMS. As before, we have smoothed the annual mean income estimates with a centred MA(9). Figure 11 is the key evidence in favour of universal character of the mean income dependence on real GDP per capita. Despite all differences in sources of data and measuring procedures the shapes of two curves separated by 20 years full of economic, social, demographic and so on processes and events in the UK as well as the country boarders and the intrinsic difference in all aspects of life between the UK and USA are practically identical! In order to highlight the level of similarity between these two curves we added two lines corresponding to 1991 and 1993 in the USA. The 1991 curve deviates from the UK curve, while the 1993 curve is very similar to the 1992 curve. The GDP per capita estimates are $22,875, $23,363, and $23,690 in 1991, 1992, and 1993, respectively. Unfortunately, we do not have any IRS data for 1992. The small discrepancy in two curves in Figure 11 between 25 and 29 years of age could have the same cause as the deviation in Figure 3 – the IRS mean income for younger ages is smaller than that measured by the Census Bureau.

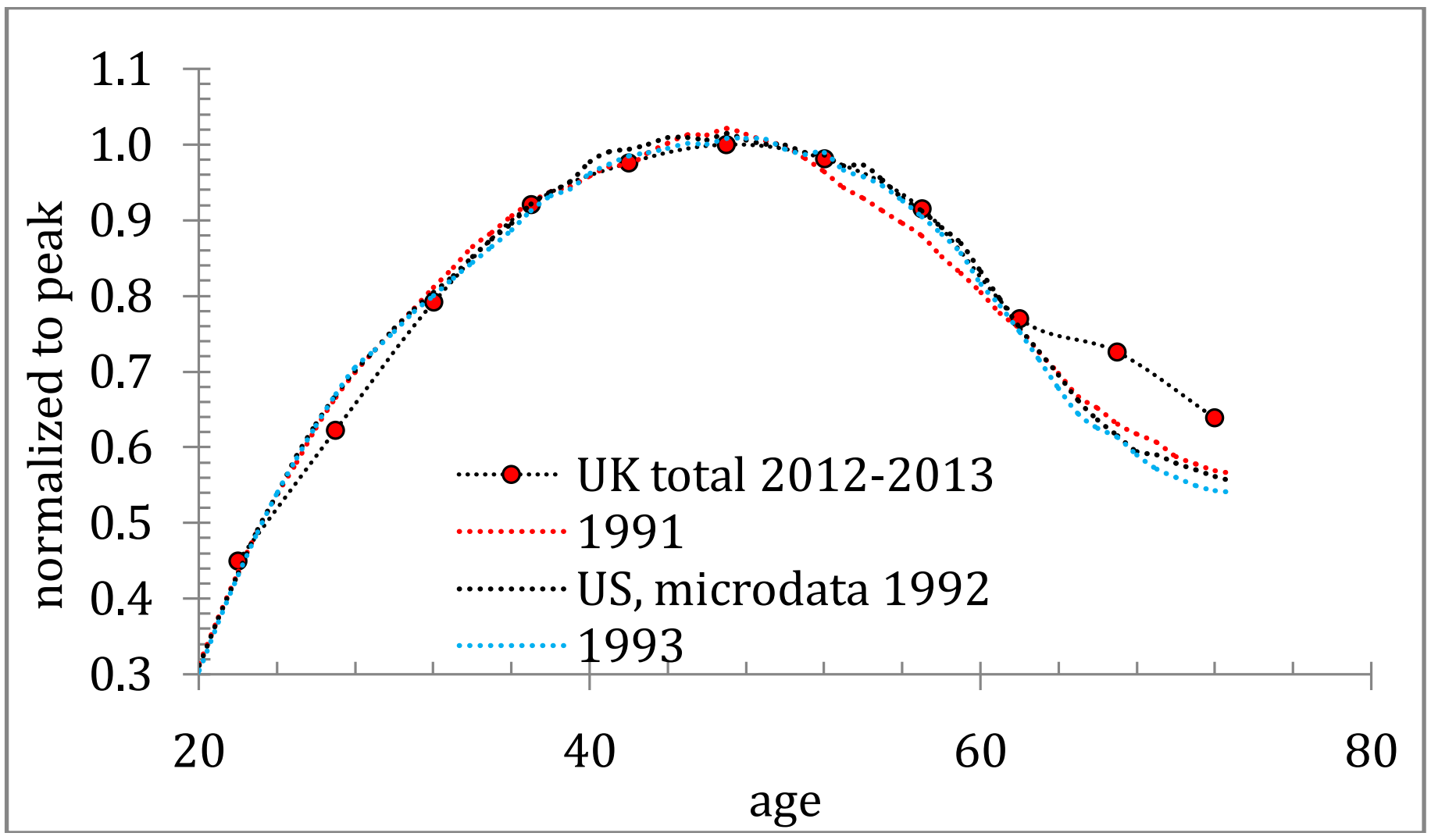

Figure 11. Comparison of the UK mean income curve for 2012-2013 and that observed in the USA in 1992. Microdata provided by the IPUMS are used.

For the earliest reliable UK mean income curve (2000-2001) we found the best matching year as well. This is the mean income curve for 1984. Figure 12 compares two corresponding UK and U.S. curves and demonstrates almost the same level of fit. The only difference – two curves deviate at ages above 55. According to the Conference Board, the level of real GDP per capita in the UK in 2000 was $20,207 (1990 US dollars). Almost the same figure was observed in the USA in 1984 - $20,122. Here, we have to stress that the time delay in real

GDP per capita between two compared countries plays no role for mean income distributions. Time is just a parameter useful for indexing GDP measurements.

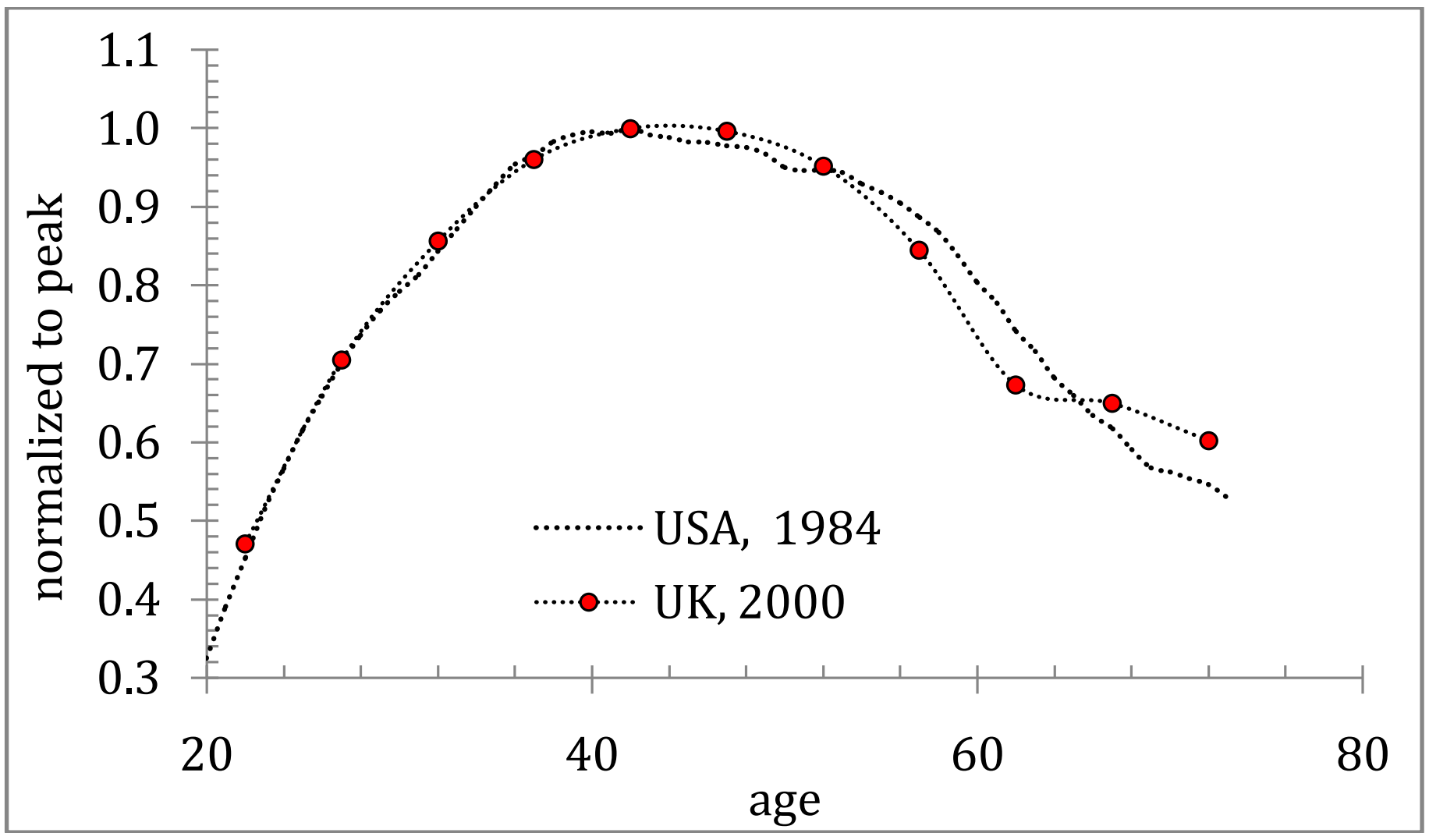

Figure 12. Comparison of the UK mean income curve for 2000-2001 and that observed in the USA in 1984.

For the UK, we have analysed a relatively short time series of the distribution of mean (tax-related) income with age and found matching distributions in the much longer CPS income dataset available for the USA. The most important result of our analysis is the universal functional dependence of the aggregated income characteristics on real GDP per capita. The UK GDP per capita is lagging by a few thousand USD (in 2014, approximately $9000 as expressed in 1990 USD) behind that for the USA. The shape of the current mean income distribution UK repeats that observed in the earlier 1990s in the USA. This means that one can project the evolution of the mean income distribution in the UK by $9000 ahead. This makes 15 to 25 years depending on the annual growth in real GDP per capita. Having the projection of major features in the personal income distribution one can develop a wise socio-economic policy to mitigate the most damaging effects. Projection beyond the current distribution in the USA is also possible since the evolution of mean income is driven by real GDP per capita only.

The UK does not provide the distribution of people over income and we could not make any estimation and comparison of the portion of people with the highest incomes. It is the second important indicator we use to characterize the increase in the age when population of developed countries achieves the peak income. This portion is most sensitive to age and real GDP per capita, especially in the first few years of work experience. For the USA, we have already demonstrated these effects [2].

### 2.3. New Zealand

Statistics New Zealand provides information on individual and household income, including wages and salaries, self-employment, government transfers, and investment income. Here, we analyse only the age dependence of mean income [7]. All income estimates are collected in the New Zealand Income Surveys (NZIS), which is an annual supplement to the Household Labour Force Survey (HLFS), conducted during the June quarter (1 April to 30 June). Based on NZIS data a comprehensive range of income statistics is produced. The corresponding tables for the years between 1998 and 2014 were downloaded from the NZIS website. They cover almost the same period as in the UK. Among other estimates, these tables contain individual weekly mean incomes in relatively narrow (5-year-wide) age groups.

There exists one potentially significant problem with the NZ income survey. The HLFS includes only 15,000 households, randomly selected throughout New Zealand, and the final NZIS personal income dataset consists of approximately 28,000 records. This is definitely not enough to cover the whole diversity of incomes depending on age, gender, and ethnicity. In the USA, the CPS includes more than 80,000 households and more than 200,000 individual records. Nevertheless, the CPS PIDs demonstrate measurable problems with the higher income estimates, which are subject to fluctuations between adjacent income bins and with time. Therefore, one may suggest that the New Zealand data demonstrate even larger fluctuations in the aggregate income estimates between adjacent age bins than we have observed in the CPS data. In this context, we consider the term "fluctuations" as the deviations from smooth curves described by a saturation function 1-exp($-\alpha/t$), where $t$ is the work experience and $\alpha>0$, before the critical age and by exponential fall above this age [1].

Figure 13 displays seventeen (nominal) mean income curves expressed in NZD, which illustrate the evolution of the age-dependent mean income in New Zealand from 1998 to 2014. Since we use nominal incomes, the earliest (1998) curve (black dotted line) is below all other curves almost everywhere, except may be the smallest work experience. The midpoints of all 5-year bins are shown by circles. The 2014 curve is above others almost everywhere. As was discussed above, the deviation from theoretically predicted smooth lines, similar to those in Figure 3, is relatively high and thus may bias the comparison with the U.S. The largest problem may arise for work experience above 40 years, where some curves (*e.g.*, 2002, 2007, 2010, and 2013) demonstrate the largest deviation from the adjacent curves. We leave the explanation of these unexpected features with Statistics New Zealand.

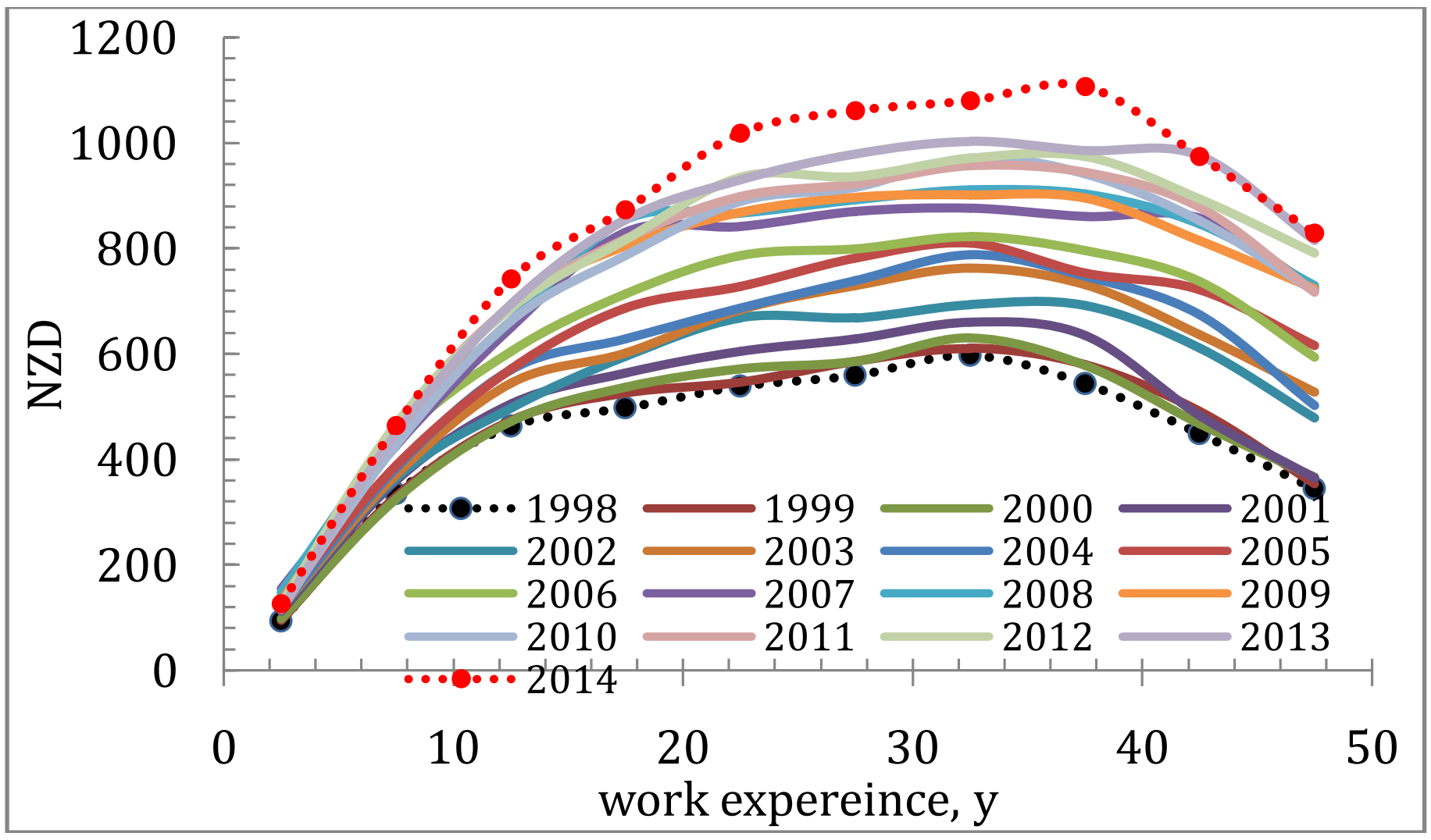

Figure 13. The evolution of the age-dependent mean income in New Zealand from 1998 to 2014. The 1998 curve (black dotted line) is below all other lines almost everywhere. The midpoints of all 5-year bins are shown by circles. The 2014 curve is above others almost everywhere. The deviation from the expected smooth lines, like those in Figure 3, is relatively high and may introduce some bias in the comparison with the US.

As discussed in paragraph 2.2, one way to carry out a cross country comparison of mean incomes is to represent them in normalized form. In Figure 14, we display the curves from Figure 13, but all divided by their respective peak values. As for the UK, there are two important observations – the work experience corresponding to the peak mean income increases with real GDP per capita and the 2014 curve now lies below all other curves before the peak value and above almost all curves, except few, after the peak work experience. In other words, the critical age increases with time while the relative mean income of the younger population decreases.

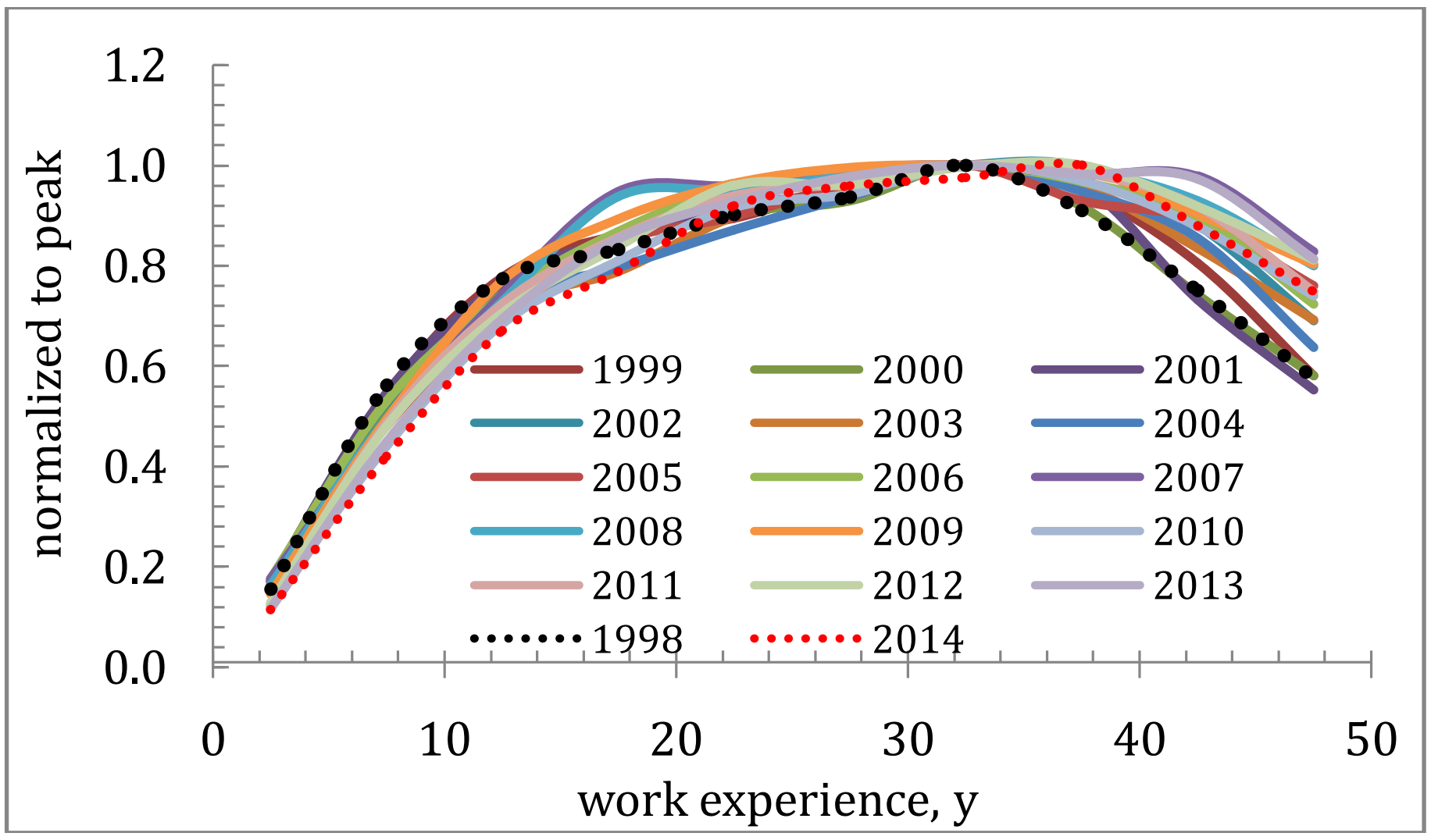

Figure 14. Same curves as in Figure 12 with all mean income estimates normalized to the peak mean incomes for the respective years.

Using the central segments of the dimensionless curves shown in Figure 15 we estimate the age of peak mean incomes. For the 1998 curve, we find the peak slightly above 32 years of work experience. Similar estimates are made for the years between 1999 and 2001. Unexpectedly, the 2002 curve peaks at 35 years. Neglecting this and other years of anomalous behaviour, we observe that the peak robustly shifts from 32 years to 36-37 years of work experience in the period from 2009 to 2014. Therefore, the observed change in the age of peak mean income is about 5 years.

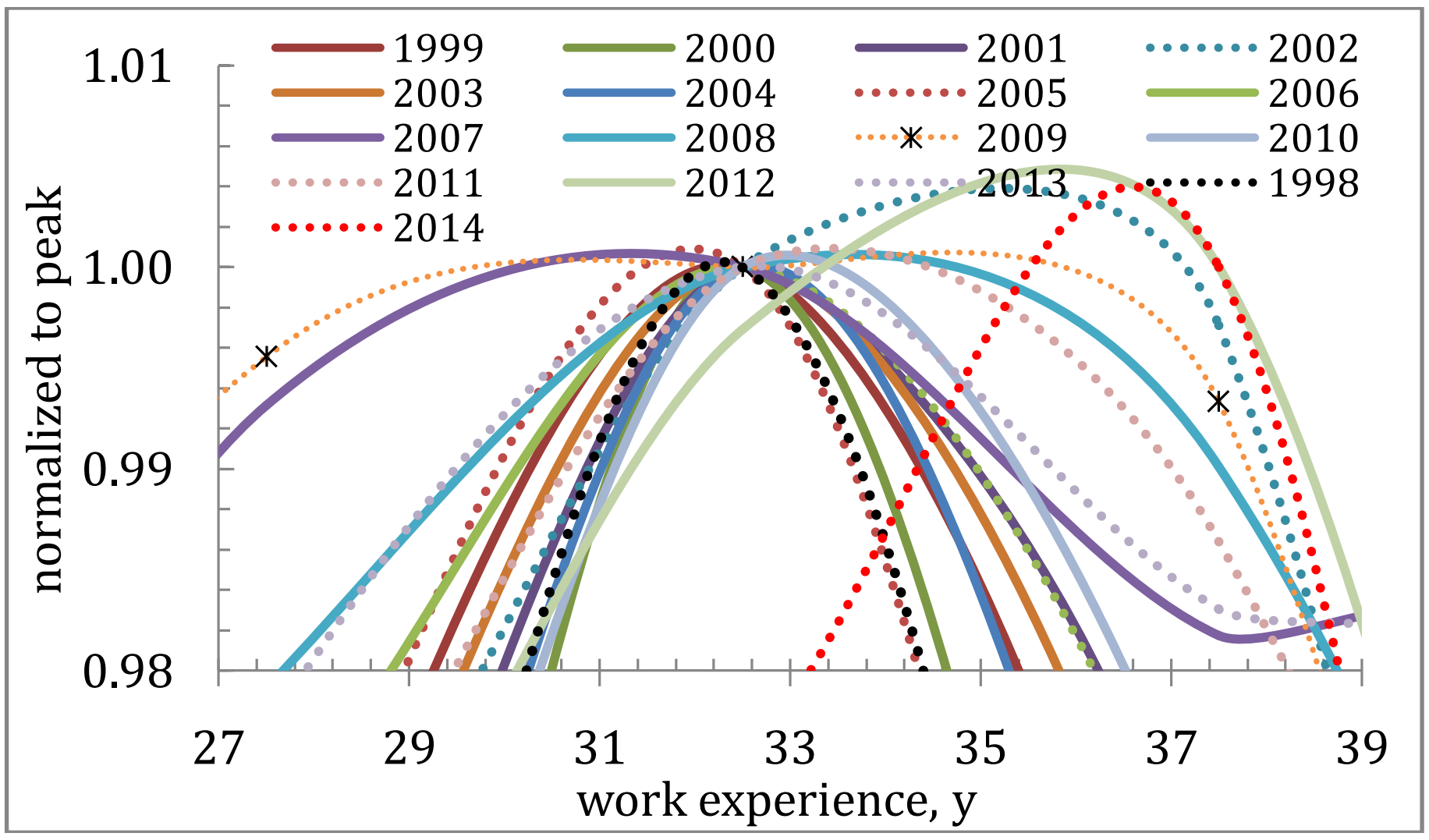

Figure 15. The central segments of the curves in Figure 14 are used to estimate the change in the age (work experience) of peak mean income.

The estimates of real GDP per capita in New Zealand are $15,404 and $20,526 for 1998 and 2014, respectively. Theoretically, the working experience should increase from 32 years to 32√(20526/15404)=36.9 years. The difference between theoretical estimates for 1998 and 2014 is ~5 years. It is in excellent agreement with the observed change. The square root dependence of the peak age on GDP per capita is well confirmed by the evolution of mean income in New Zealand.

Despite the fact that the theoretically predicted change in the critical age confirms the observed change, there is a significant discrepancy with the UK peak age estimates. The level of real GDP per capita in New Zealand lags behind that in the UK by a few thousand USD. In 1998, this difference was $3,623 and then slightly fell to $3,273 in 2014. Under our framework, the peak age in the UK has to be larger than that in New Zealand by several years, but we observe an opposite situation. This contradiction needs detailed investigation, but here we just refer to the accuracy of GDP estimates based on PPP conversion. There are some controversial estimates, which indicate significant problems for some countries. In 2014, TED provides the following estimates (in 1990 US$) of real GDP per capita: Spain - $16,217, Portugal - $13,429, Estonia - $22,665, Belorussia - $15,144. Hence, there are some doubts in the accuracy of TED readings for some countries.

On the other hand, the evolution of mean income with GDP provides an excellent and independent tool to measure the relative level of GDP. If to consider the US GDP as a reference, New Zealand peak age of 36 years in 2014 corresponds to real GDP per capita observed in the USA in 1996. This makes ~$25,000. In turn, the level in the UK in 2011 (32 years of work experience) corresponds to 1983/1984 - ~$19,500. These are only crude estimates obtained from the peak age, but they definitely highlight some problems with the GDP estimated reported by TED. Much more accurate estimates might be obtained by matching the whole mean income curves.

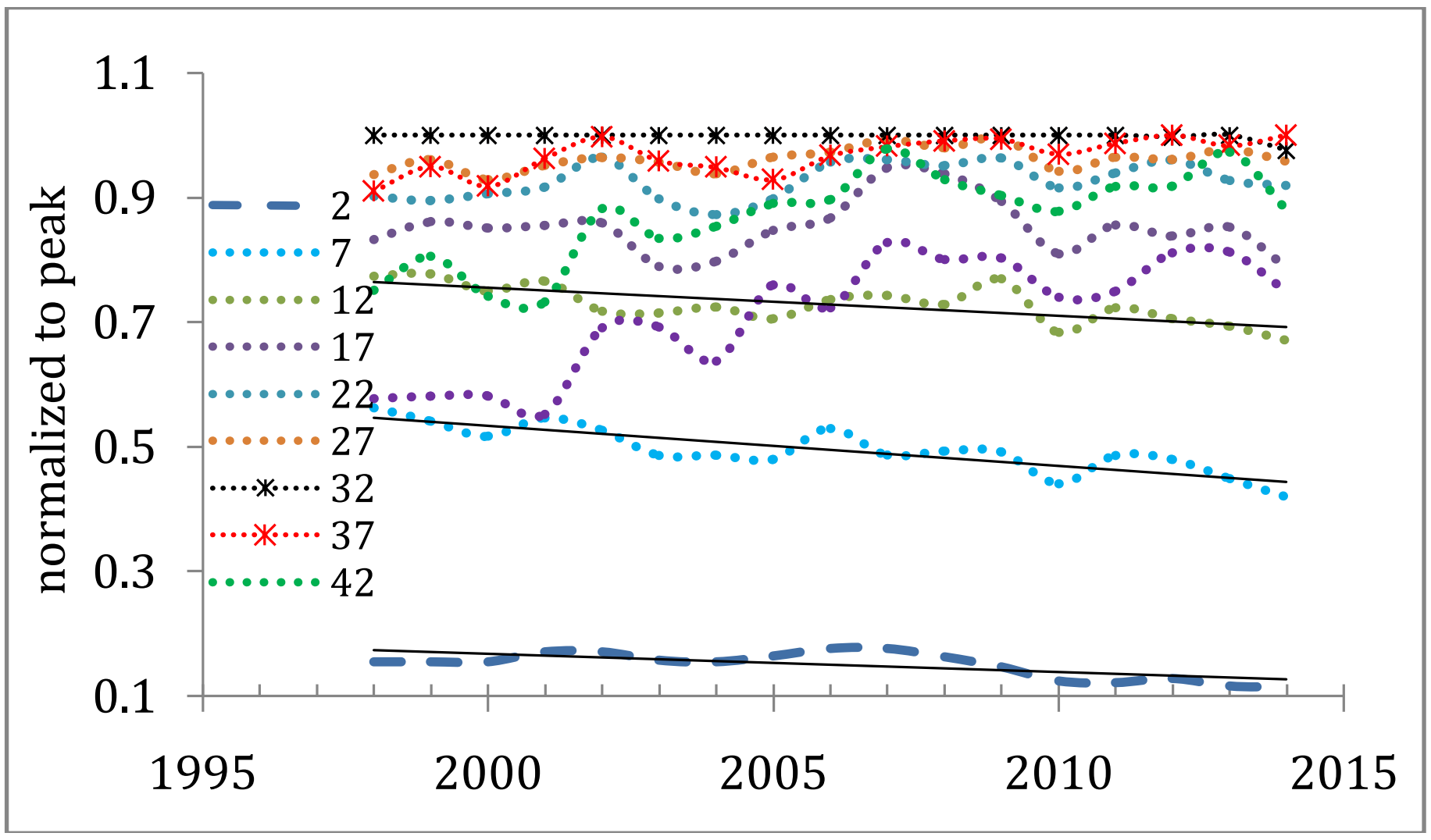

Figure 16. The evolution of mean income in all 5-year age groups normalized to the peak value in the same year. In the youngest age groups (0 to 4, 5 to 9, 10 to 14, and 15 to 19 years of work experience), the portion of mean income has been falling since 1999. The peak work experience shifted from the 30 to 34 years bin to the 35 to 39 years bin in 2014.

Figure 16 is similar to Figure 10 for the UK and displays the evolution of peak-normalized mean income in various age groups. In line with the observations in Figure 15 and the above discussion, the peak work experience resides in the bin from 30 to 34 years. The peak jumps into the elder group in 2012, and then in 2014. The mean income in the group “32” started to fall relative to that in the group “37”. The portion of mean income has been increasing in all elder groups and decreasing in the younger groups. Figure 17 elaborates on the trends observed in the age bins around the peak value. From this Figure, the peak age will likely move into the next age group at a horizon of 10 to 15 years. Overall, in the USA, UK and New Zealand we observe robust indications of a gradual increase in the age of peak mean income, which shifts from younger age bins to elder bins. This is a consequence of increasing GDP per capita.

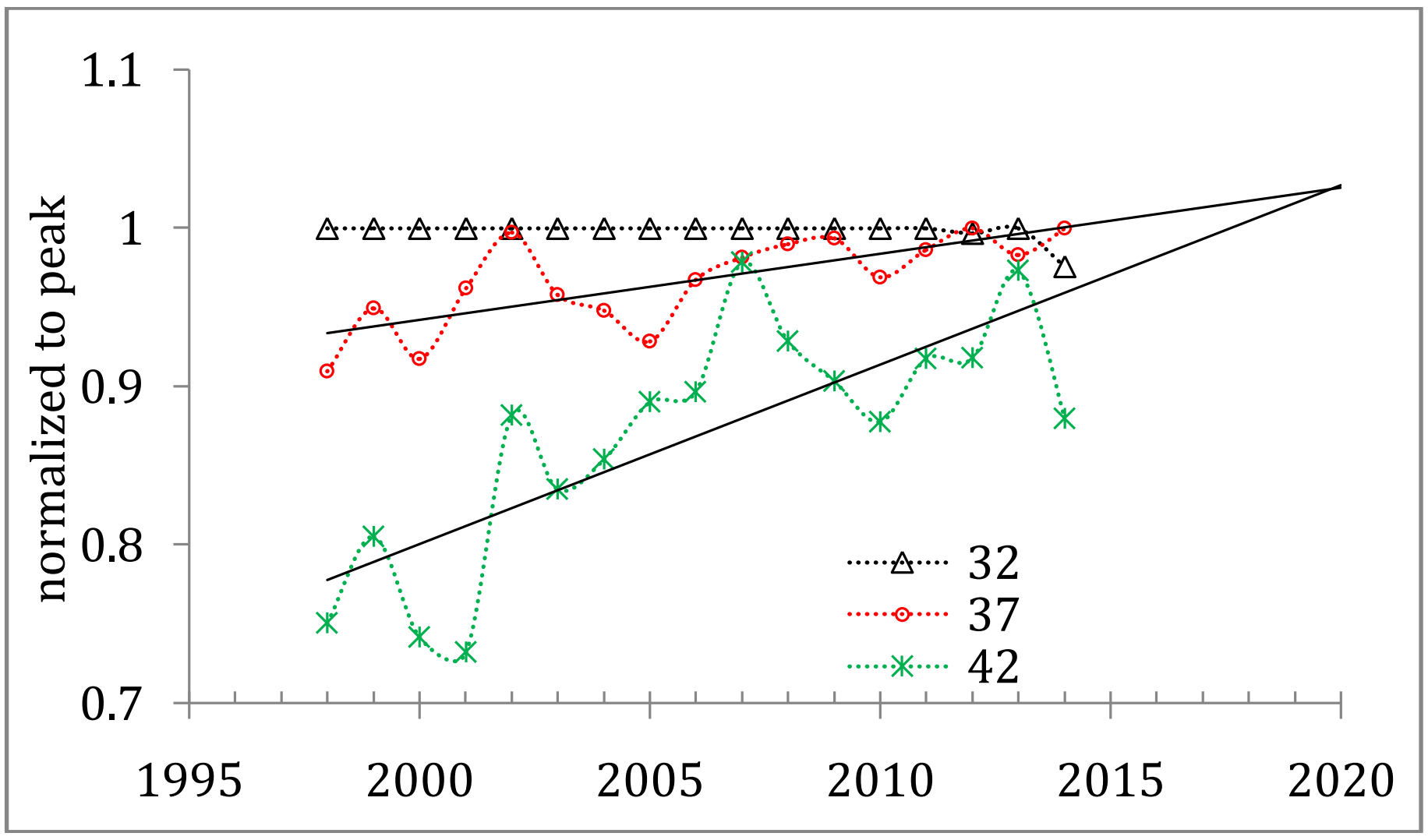

Figure 17. Trends in the evolution of the relative mean income in three age groups.

Now, we have come to the most sensitive method of cross country comparison – fitting the normalized mean income curves with time delays of decades. The results of this method best illustrate the level of match in income distribution between the studied countries. As suggested in paragraph 2.1, the NZ curve for 2014 has to match the U.S. curve for 1985. However, the peak mean income for the NZ2014 curve better corresponds to 1996. So, the best fit between the U.S. and NZ curves should be the decider. As for the UK, we use the set of income microdata published by the IPUMS. The original mean income estimates in one year age intervals are smoothed with a centred MA(9).

Figure 18 presents three panels with New Zealand and matching U.S. curves for 2014, 2006 (the midpoint of the 1998-2014 interval) and 1998. The best-fit curves in the U.S. are 1988 ($20,525 in NZ vs. $22,500 in the U.S.), 1980 ($19,028 vs. $18,577) and ($15,403 vs. $15,304). Overall, the fit between the NZ2014 and US1988 is excellent except the period between 25 and 39 years of work experience. This deviation might be the reason of the overestimated critical age in Figure 15. One of possible causes for the observed fluctuations near the peak mean income is the poor representation of the very high-income individuals in the NZIS. They could be just missing from the survey because the sample is sparse. A similar effect is observed with the youngest black women in the USA. For the NZ2006 curve, the overall fit by the US1980 curve is even better than that for the NZ2014. In 2006, real GDP per capita in New Zealand was by $500 larger than that in the USA in 1980. This might indicate some problems in the related GDP estimates in New Zealand as well as larger fluctuations in the shape of mean income curves measured in the U.S. and New Zealand. For the NZ1998, the overall fit is the worst, as might be expected from lower accuracy of income measurements in the past. Taking into account all fluctuations in GDP and income measurements in New Zealand and U.S. the overall match and the evolution of shape both support the universal character of the mean income dependence on real GDP per capita.

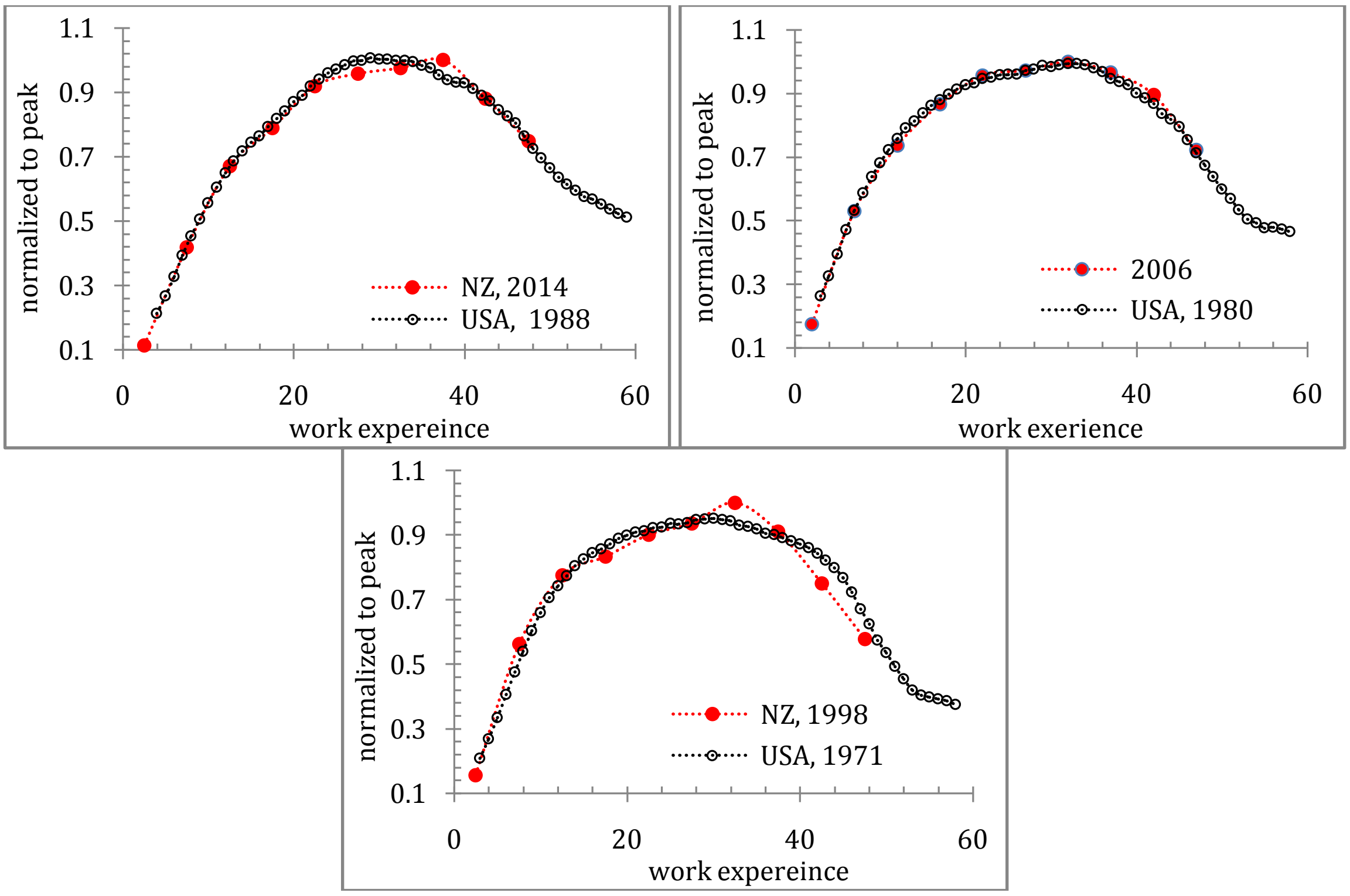

Figure 18. Comparison of the normalized mean income (weekly income from all sources) curves measured in New Zealand USA.

New Zealand provides a slightly longer time series of mean income, which covers a wider range of real GDP per capita than in the UK. For New Zealand, the ratio of GDP per capita in 2014 and 1998 is 1.33. For the UK, the ratio of GDP estimates in 2011 and 1999 is only 1.18. For the total growth in the critical age during the corresponding periods these estimates give factors 1.15 and 1.08, respectively. The set of income data provided by New Zealand might be superior if not the observed level of fluctuations in income estimates and the doubts in the PPP conversion of NZD into USD. The lesson learned with the NZ data is that income measurements have to be carefully conducted and all problems should be resolved before reporting them to the broader scientific community. It is difficult to believe that the observed fluctuations are real.

### 2.4. Canada

We finish our country analysis with Canada, which provides an extended set of income data. These data cover all persons who completed specific tax return forms for the year of reference. Therefore, the data are related to tax purposes only and some non-taxable income sources are missing from the dataset. In 2013, about 74.9% of Canadians (of all ages) filed tax returns. Most children do not file tax return as well some elder people. That may introduce age-specific bias in the estimates for children and pensioners. Unlike the CPS survey, all data are extracted from administrative files. The sample includes 100% of individuals who filed an individual tax return but not all population. Depending on age, from 89% to 96% of the population is covered by administrative records [8]. Overall, the population and income source coverage is good for the purposes of our research.

Figure 19 depicts the evolution of real mean income since 1976 [9]. Three years age shown for comparison – 1976, 1993 and 2011, i.e. with approximately a 17 year step. The real mean income estimates are given in 2011 constant Canadian dollars and calculated in 10-year bins except the two youngest age groups: under 20 years – the average age of 17 years is used, and between 20 and 24 years of age. The curve for 2012 presents real mean incomes only in 10-year bins; it was obtained from a different table provided by Statistics Canada.

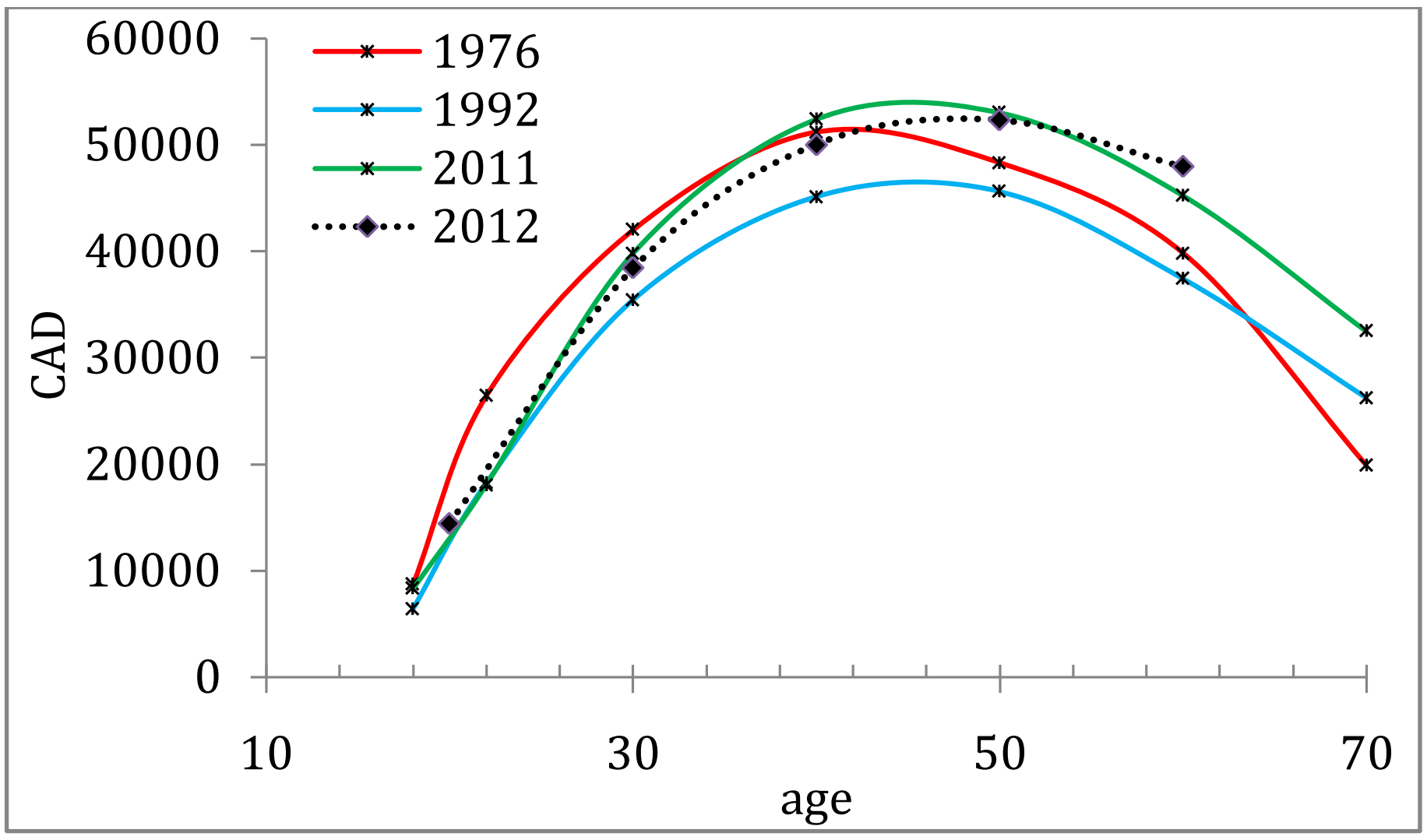

Figure 19. The evolution of age-dependent real mean income (in 2011 constant Canadian dollars) in Canada from 1976 to 2012. All estimates are given in 10-year bins except the two youngest age groups (below 20, and between 20 and 24). The curve for 2012 presents mean incomes only in 10-year bins and was obtained from a different table provided by Statistics Canada. The highly unusual feature is the 1976 curve above the 1993 curve.

The first and highly unusual observation is that the 1976 curve is above the 1993 curve. The real GDP per capita curve in Figure 5 gives $18,138 (1990 US$) in 1992 and $14,902 in 1976. For comparison, the mean income for the whole population with income in Canada is presented in Figure 20. In fact, the average income in 1976 was $35,700 and only $33,300 in 1993. The contradiction between real GDP per capita and mean income is likely related to the difference between domestic currency and that converted at Geary Khamis PPPs. In our quantitative analysis, this discrepancy may introduce large errors in theoretical estimates of the peak age.

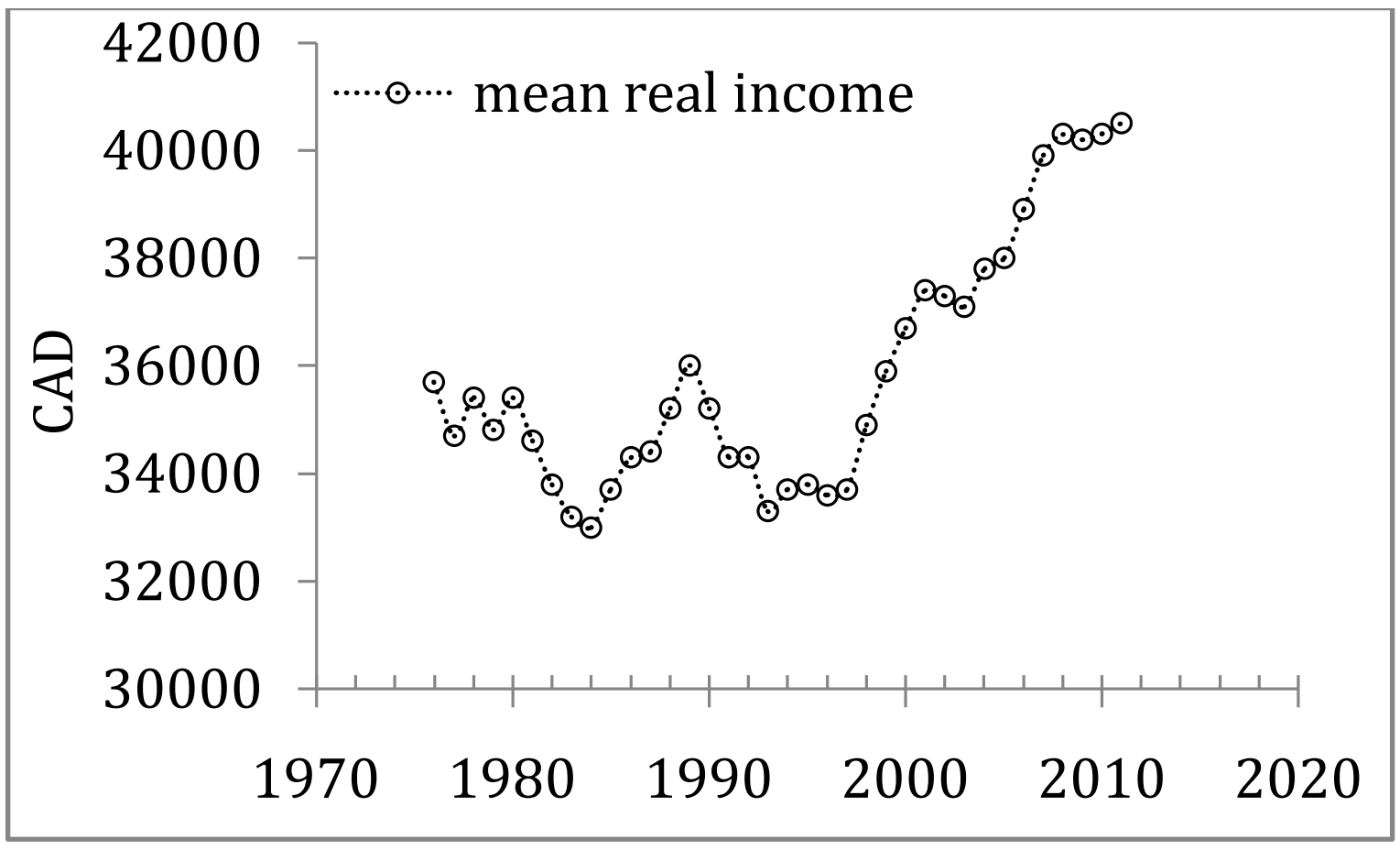

Figure 20. The evolution of mean income (2011 CAD) in Canada between 1976 and 2011.

Following the procedure developed in two previous paragraphs, we normalize the curves in Figure 19 to their respective peak value and present the central segments of the obtained curves in Figure 21. The 1976 curve peaks between 41 and 42 years of age (27 to 28 years of work experience). In 1992, the peak age is about 45 years, and then it reaches 49 to 50 years in 2012. Overall, the age of peak mean income increases by about 8 years. Theoretically, the change in real GDP per capita from $14,902 to $25,629 should change by a factor of 1.31, i.e. the peak working age experience rises from 27.5 years to 36.1 years (50 years of age). This is an extremely accurate prediction, especially taking into account the problems with real GDP per capita estimates. The proportion of working age population does not change much in the USA since the mid-70s. It is likely not a big error if we assume that the share of working age population in three studied countries does not change since 1976. Then the correction for population does not affect the relative change in real GDP per capita and the estimates of the peak age increase in Canada, New Zealand and the UK are not biased.

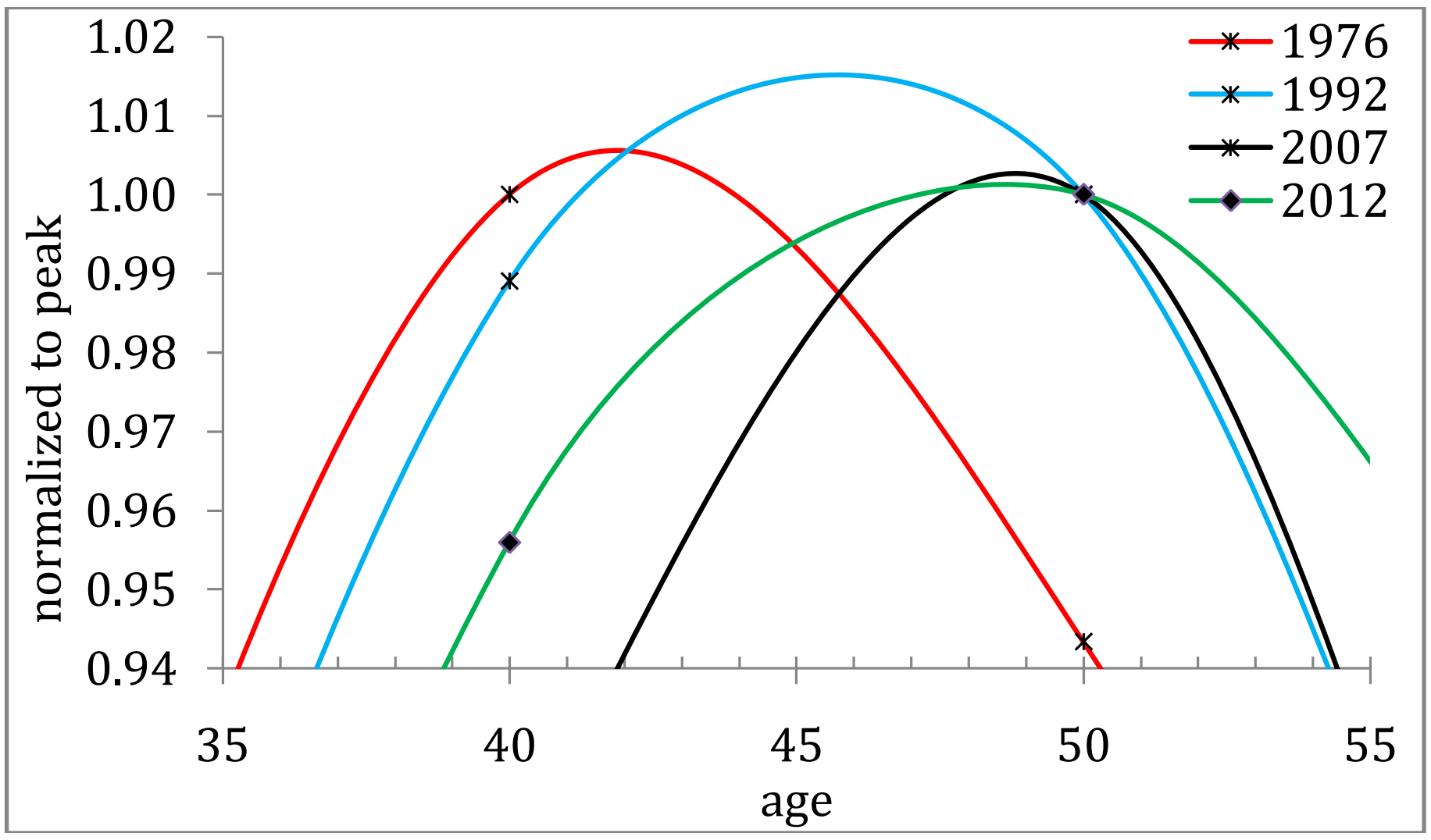

Figure 21. Three curves in Figure 19 normalized to their respective peak values. The age of peak mean income value increases with time from 42 years in 1976 to 49 years in 2012.

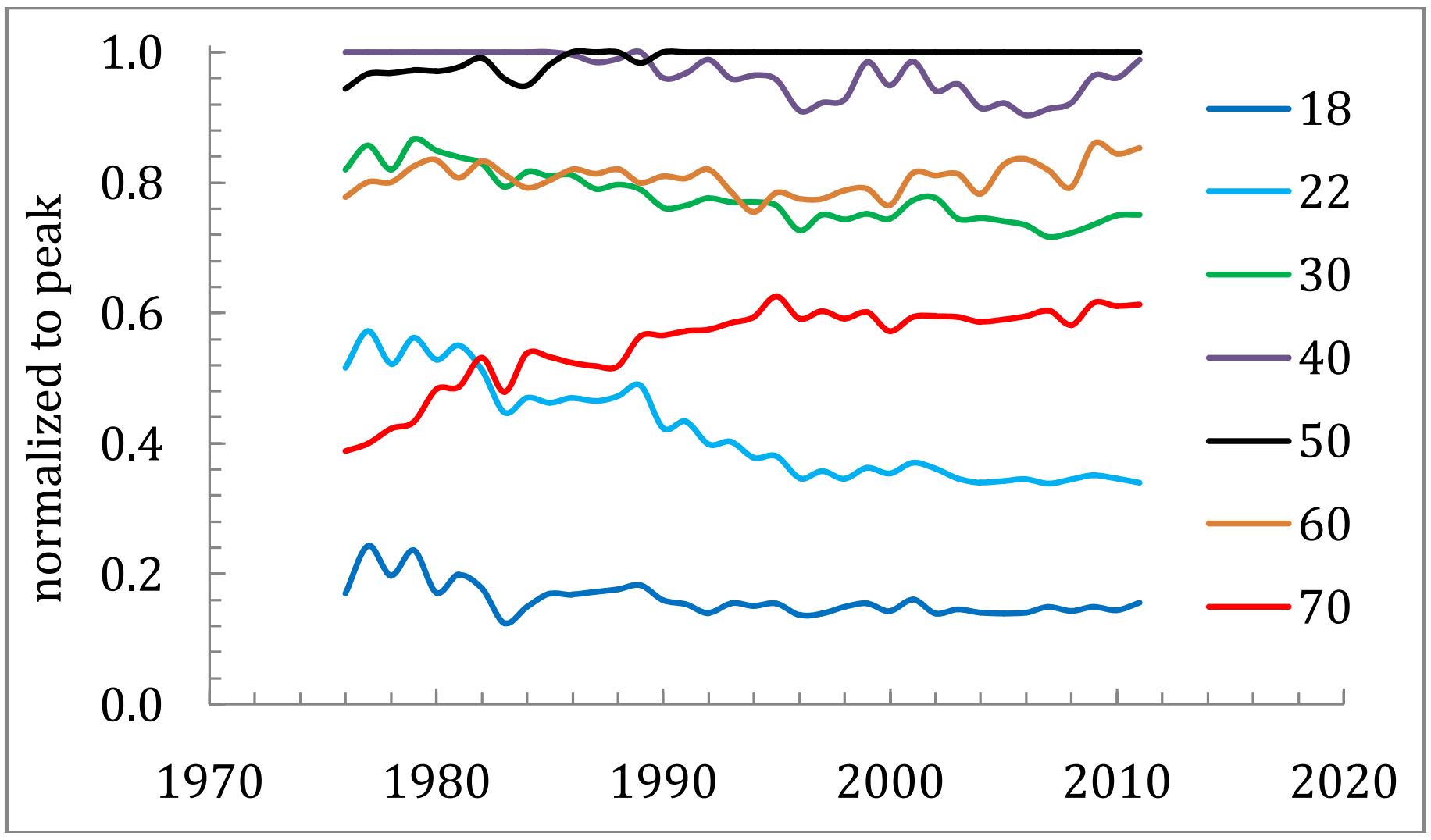

Figure 22. The evolution of mean income in all 10-year age groups normalized to the peak value in the same year. In the younger age groups (“18”, “22”, and ”30”) the proportion of mean income has been falling since 1976.

Figure 22 is similar to Figures 10 and 16 and displays the evolution of peak-normalized mean income in various age groups. The peak age resides in the bin from 35 to 44 years of age before 1990. Then the peak jumps into the elder group and will likely stay below 54 years before the group between 55 and 64 overtakes the lead. As in other three countries, the proportion of mean income has been increasing in all elder groups and decreasing in the younger groups. In Canada, fluctuation of the curves in Figure 22 is much lower than that for New Zealand and likely at the same level as in the UK. This is a consequence of income data quality – better coverage of population is directly translated into the accuracy of mean income estimates.

Figure 23 depicts two panels where the normalized mean income curves observed Canada in 2011 ($22,994) and 1980 ($12,931) are matched by U.S. curves for 1987 ($21,788) and 1962 ($11,904), respectively. The excellent match between Canada2011 and USA1987 curves well corresponds to GDP estimates, while the GDP levels for 1980 in Canada and 1960 in the USA differ more in relative terms. As we discussed above, the GDP values for the USA before 1975 have to be corrected for the effect of changing working age population. Between 1987 and 1962 the correction is approximately 7%. This makes the 1962 estimate to increase to $12,708. So, Canada provides the longest time series of mean income among three studied countries with the largest factor of peak age increase - 1.31. This makes our measurements more precise and allows much better fit between observed and predicted change in the peak mean income.

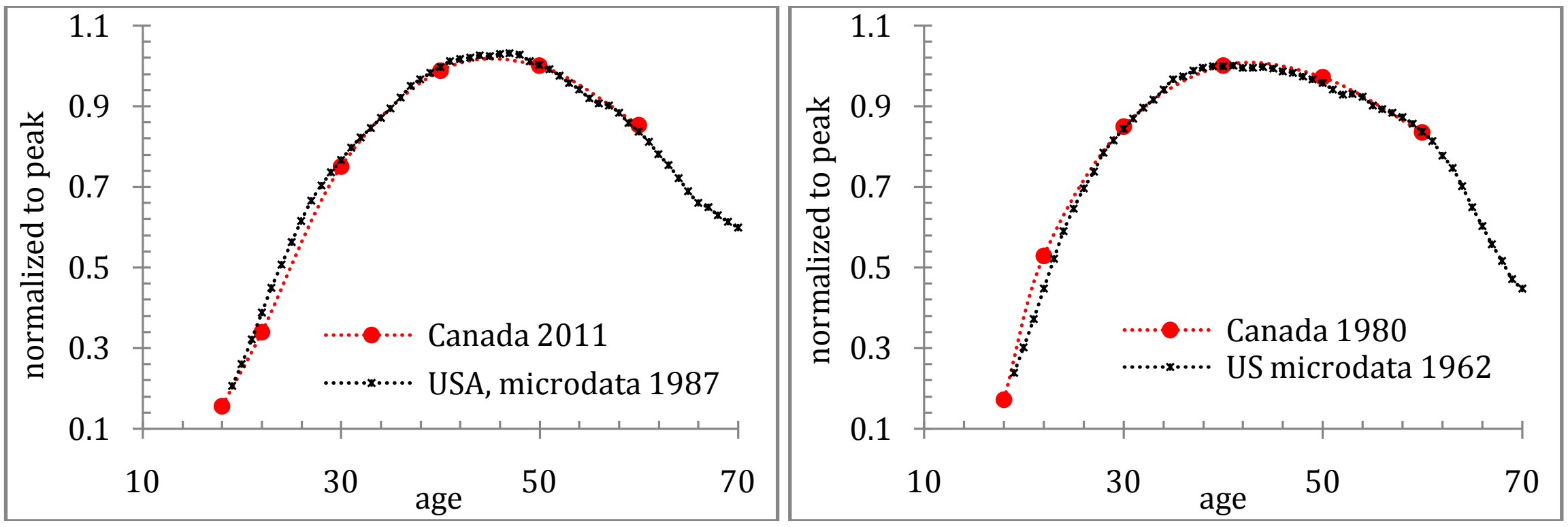

Figure 23. The evolution of mean income in Canada in 2011 and 1980 is matched by the curves measured in the USA in 1987 and 1962, respectively. (No income microdata are available before 1962.) Age bins are 10 years for Canada. For the USA, we use microeconomic data with annual mean income smoothed with a MA(9).

Canada is the only country from the studied trio reporting age-dependent PIDs for the higher incomes. Figure 24 displays the number of people with income above a given income threshold as a function of age. We have selected different thresholds to retain the portion of population: $100,000 in 2000, $100,000 in 2006, and $150,000 in 2013. All age bins are 10 years, except the youngest between 0 and 24 years of age. The youngest bin is prone to strong bias because it includes fluctuating numbers of children with incomes. The 2006 curve is higher than the 2000 curve because they have the same threshold but the total nominal income in 2006 is much larger than in 2000, and thus, more people have larger incomes. Moreover, the population pyramid changing with time may introduce a significant bias into the number of people of a given age. In Figure 24, three curves peak at different ages. To suppress the population effect we have scaled the portion of people with the highest income to the total population with income in the same age bin. Figure 25 shows three normalized curves. As one can see, the age pyramid effect is removed and all curves peak at the same

age. Summing the number of people above the threshold in all age groups and dividing it by the total population with income, we calculate the total portion of people with incomes above the threshold. For the thresholds in Figure 24, this portion is 2.5% in 2000, 4.5% in 2006, and 2.8% in 2013.

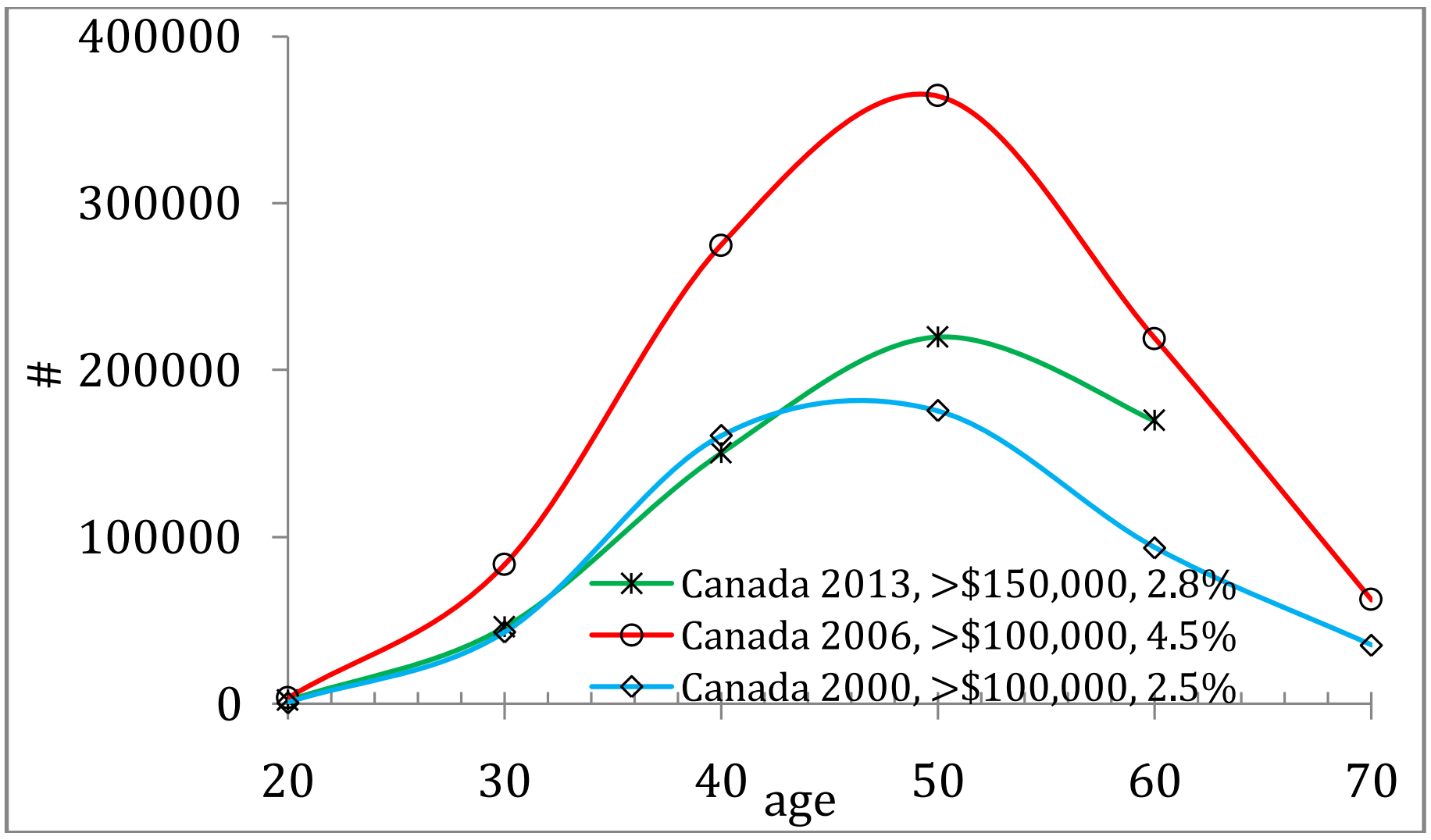

Figure 24. The number of people with income above a given income threshold as a function of age. Thresholds are $100,000 in 2000, $100,000 in 2006 and $150,000 in 2013. Age bins are 10 years, except the youngest between 0 and 24 years of age. The number of people depends on threshold and population in each bin.

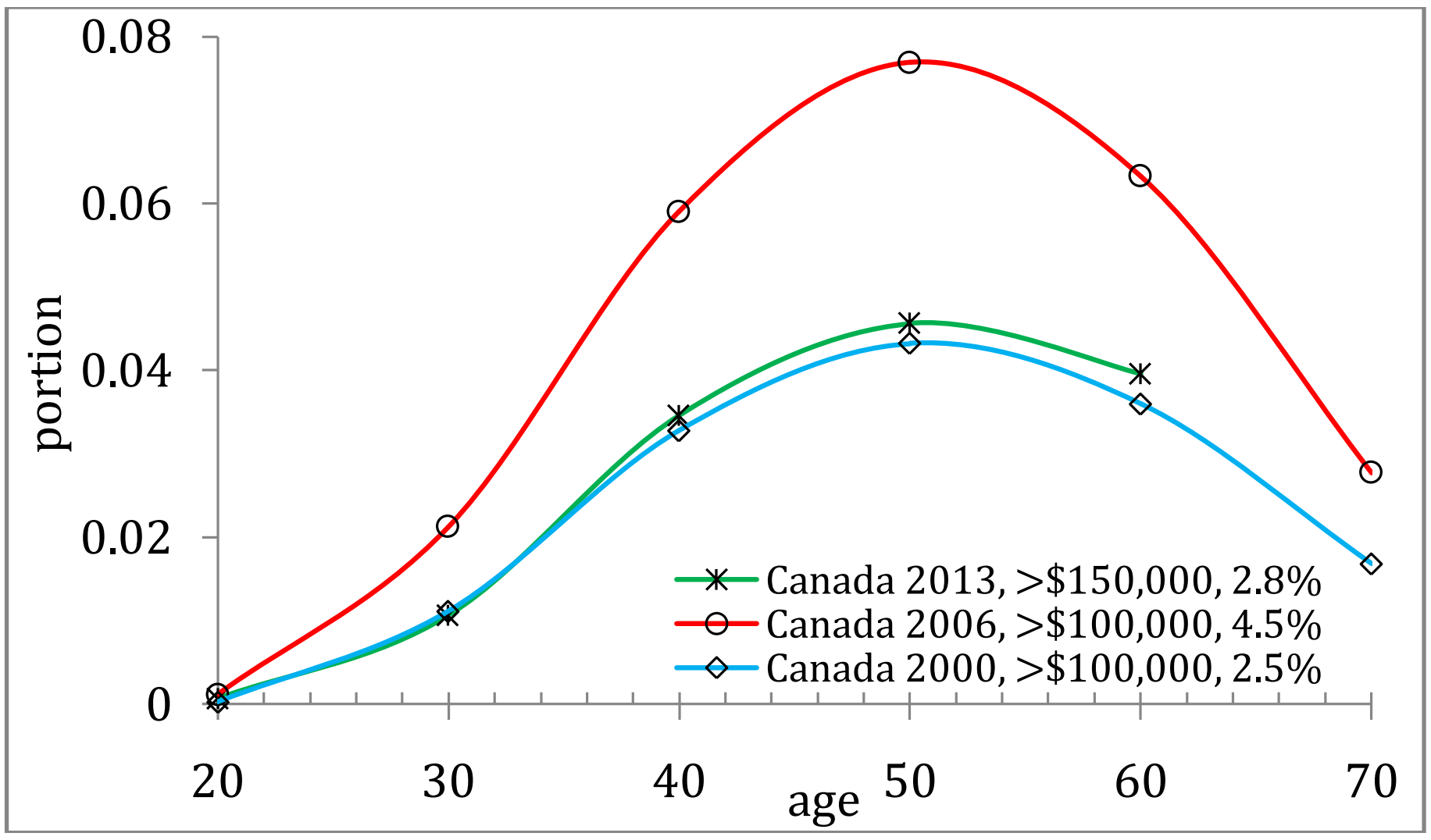

Figure 25. The portion of people with income above a given income threshold as a function of age: the number of people above the threshold in a given age bin is divided by the total number of people in the same bin. The age pyramid effect is suppressed. The cumulative portion of people above the threshold is 2.5% in 2000, 4.5% in 2006, and 2.8% in 2013. Notice the same bin and threshold settings as in Figure 24.

When the total income and population are subject to significant changes with time the best way to compare the age-dependent income distributions in different years is to normalize them to their respective peak values. Then direct comparison is possible, which shows the relative rate of income/population change with age. Figure 26 depicts three peak-normalized curves from Figure 25. These normalized curves practically coincide with just small differences in the age groups 25 to 34 years and 55 to 64 years. The age of the largest portion of people with the highest incomes resides in the bin between 45 and 54 years. Because of the width of this bin, which includes the true peak year for all curves, it is difficult to resolve any change.

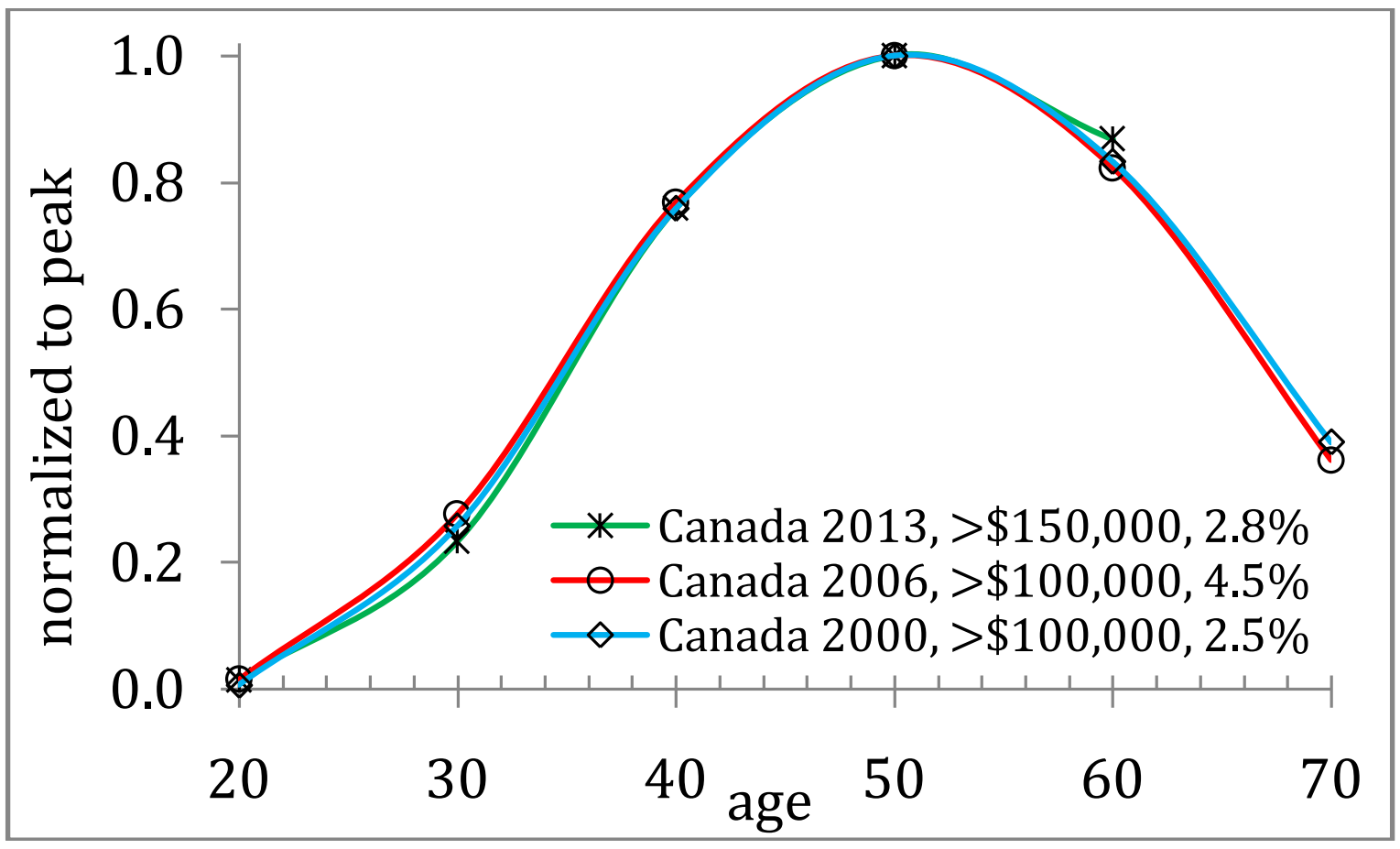

Figure 26. The curves in Figure 25 are normalized to their respective peak values. The age of largest portion of people with the highest incomes resides in the bin between 45 and 54 years. It is difficult to estimate the change in peak age. Notice the same bin and threshold settings as in Figure 24.

One problem in Figure 26 is that the 2006 curve lies above the 2000 curve, while our model and experience suggest the opposite situation. This controversy is actually related to the difference in the total portion of people with the highest incomes: 2.5% in 2000 and 4.5% in 2006. All PIDs available for Canada between 2000 and 2013 are characterized by $50,000-wide income bins above $100,000. The choice of threshold is limited to the boundaries of these bins. The year of 2000 is characterized by the lowermost number of people and gross nominal income, and thus, by the lowermost threshold of the Pareto distribution. We use it to illustrate the change in the age-dependent portion of people with the highest incomes for several thresholds between $50,000 and $250,000. Figure 27 illustrates the change in the age-dependent curve. The peak age grows with the threshold, i.e. more and more time is needed to reach higher incomes. This observation puts an important constraint on the direct cross comparison of the age-dependent portion of people with the highest incomes. One should use thresholds, which retain the total portion of people at the same level.

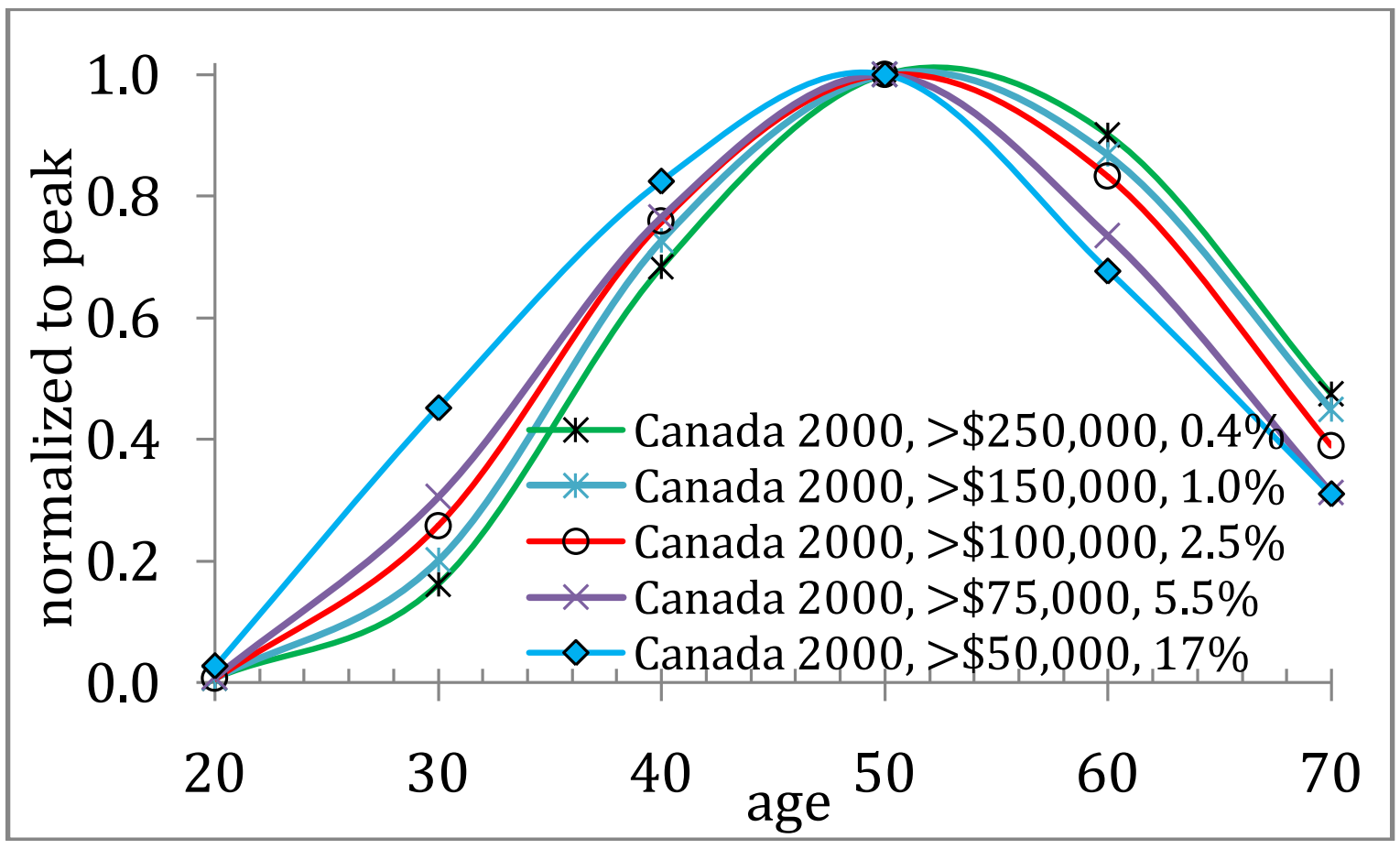

Figure 27. Canada 2000. The age-dependent portion of people with incomes above five thresholds - from $50,000 to $250,000. The peak age depends of threshold. Therefore, one has to compare curves with thresholds giving the same portion of people.

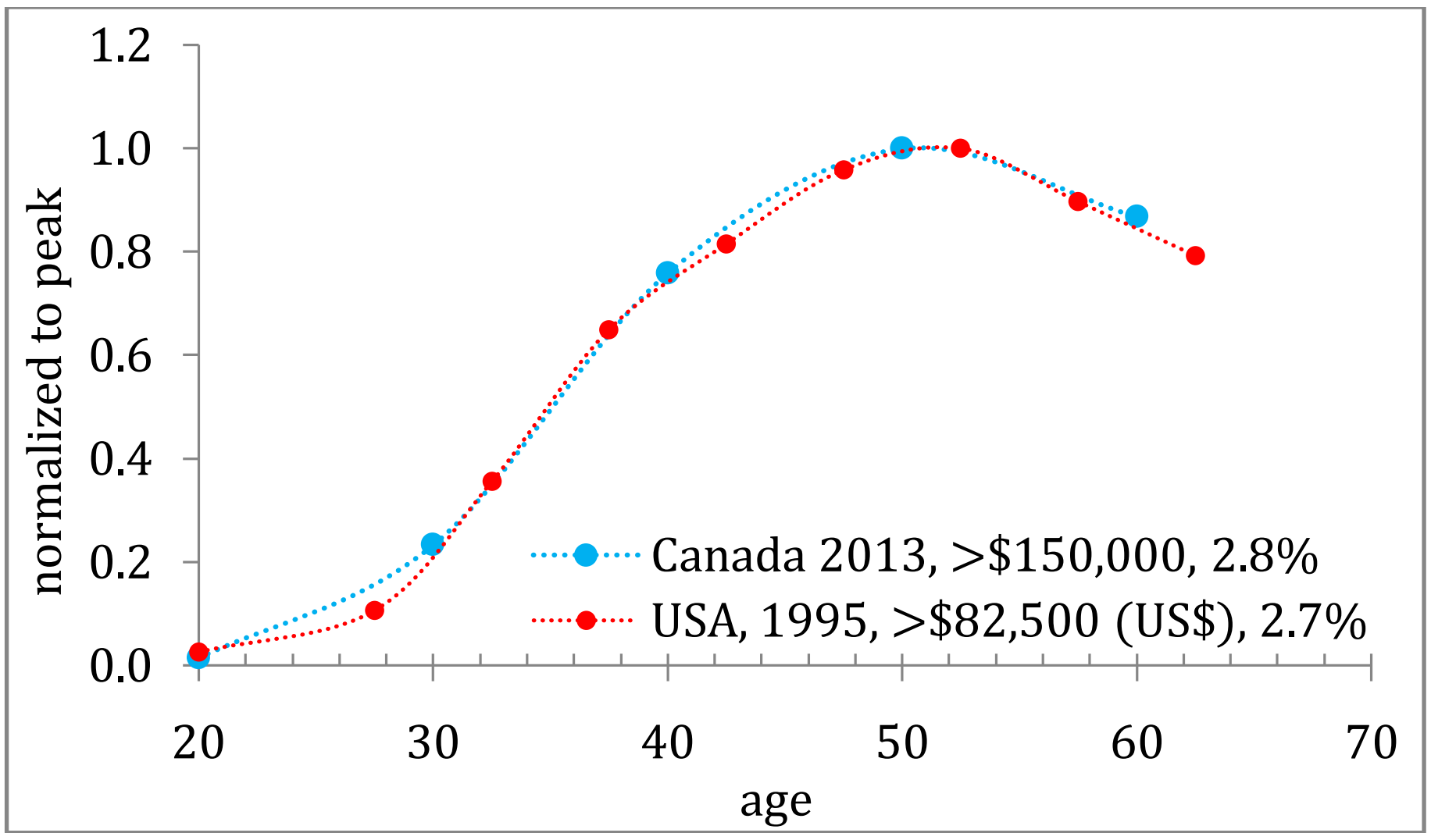

Figure 28. Comparison of the age-dependent portion of people with incomes above given threshold. The 2013 Canada curve (>$150,000, current CAD) is best fit by the 1995 US curve (>$82,500, current US$). TED reports for Canada $26,000 (1990 US$) in 2013 and $24,712 for the USA in 1995. The total portion of population above the thresholds in both cases is approximately 2.7%.

In Figure 28, we match the 2013 Canada curve and the 1995 US curve. The coincidence between the portions of total population with incomes above $150,000 (CAD) in Canada and $82,500 (USD) in the USA, both in current dollars, is striking. We retain the total portion of population above these thresholds at approximately 2.7%. That eliminates the threshold dependent bias. Total Economy Database reports $26,000 (1990 US$) for Canada in 2013 and $24,712 for the USA in 1995. The difference is not large and a 4% correction for the change in working age population makes the 1995 U.S. estimate to rise to $25,643. Hence, the fit between two curves proves that the portion of people with the highest incomes is a country-independent variable. In other words, it is a universal variable which depends only on real GDP per capita. It is important to extend the set of countries in order to support this finding.

## Conclusion

We have studied two specific features of personal income distribution in three countries: Canada, New Zealand, and the UK, and compared them with the USA. The dependence of mean income and the portion of people with the highest incomes on age are both characterized by varying length of the involved time series and their accuracy. Canada provides a set of mean income data covering practically the whole population and the period since 1976, but the data on personal income evolution with age are limited to the period between 2000 and 2013. Statistics New Zealand also reports tax-related income as obtained from a survey covering a small portion of the whole population. Higher amplitude fluctuations, likely induced by underrepresentation of the highest incomes, limit the usefulness of these estimates for the purposed of our study. In addition, there is some controversy between real GDP and income estimates reported by New Zealand and the Conference Board. The UK provides the shortest but useful time series collected by HMRC.

All data corroborate our main assumption – the evolution of income distribution is universal, follows a unique trajectory, and depends only on real GDP per capita converted at PPP exchange rates. We have proved quantitatively that the dependence of mean income on age in Canada, New Zealand and the UK as well as the age-dependent portion of people with the highest incomes in Canada do reproduce similar dependencies observed in the USA, but many years before, when the level of real GDP per capita was the same. Since the U.S. outpaces three studied countries by several thousand dollars per head, the lag reaches 20 to 25 years. For example, the mean income dependence measured in the UK in 2012-2013 one-to-one repeats that observed in the USA in 1992. The age-dependent portion of rich people in Canada in 2013 reproduces that measured in the USA in 1995. The time dependence of the studied characteristics is just parametric, however.

We have proven that the growth of work experience corresponding to the peak mean income is accurately described by the square-root function of real GDP per capita. This feature corresponds to the key assumption of our model, which predicts both studied features precisely. The coherence of theoretical predictions and long-term observations in four countries proves that the evolution of personal income (at least in these four countries) is a physical process described by a simple relationship.

The results of measurements carried out in this study well match the prediction of our microeconomic model. A unified quantitative description in four countries is useful not only from theoretical point of view as a possibility to mathematically describe the process of personal income distribution as the evolution of a physical system. The universal character of personal income evolution as a unique function of real GDP per capita allows accurate forecasting of very specific income characteristics related to fiscal, monetary and other types of socio-economic policies. Extending the observed linear trends of real GDP growth into the future (see Figure 5), one can be use the past U.S. PIDs as templates for the future PIDs in three countries at a time horizon from 15 year (Canada) to 50 years (New Zealand).